# X-ray study of krypton and xenon under pressure reveals the mechanism of martensitic transformations

A.D. Rosa[1*], A. Dewaele[2,3], G. Garbarino[1], V. Svitlyk[1], G. Morard[4], F. De Angelis[5], M. Krstulovic[6], R. Briggs[7], T. Irifune[8], O. Mathon[1] and M.A. Bouhifd[9]

1. European Synchrotron Radiation Facility (ESRF), 71, Avenue des Martyrs, Grenoble, France.

2. CEA, DAM, DIF, 91297 Arpajon Cedex, France

3. Université Paris-Saclay, CEA, Laboratoire Matière en Conditions Extrêmes, 91680 Bruyères-le-Châtel, France

4. Université Grenoble Alpes, Université Savoie Mont Blanc, CNRS, IRD, IFSTTAR, ISTerre, 38000 Grenoble, France

5. Dipartimento di Fisica, Universita' di Roma La Sapienza - Piazzale Aldo Moro 5, 00185 Roma, Italy

6. University of Potsdam, Institute of Geosciences, Karl-Liebknecht-Str. 24-25, 14476 Potsdam-Golm, Germany

7. Lawrence Livermore National Laboratory, Livermore, CA, United States of America

8. Geodynamics Research Center, Ehime University, 2-5 Bunkyo-cho, Matsuyama 790-8577, Japan

9. Laboratoire Magmas et Volcans, Université Clermont Auvergne, CNRS, IRD, OPGC, F-63000 Clermont-Ferrand, France

*corresponding author: angelika.rosa@esrf.fr

# Abstract

The martensitic transformation is a fundamental physical phenomenon at the origin of important industrial applications. However, the underlying microscopic mechanism, which is of critical importance to explain the outstanding mechanical properties of martensitic materials, is still not fully understood. This is because for most martensitic materials the transformation is a fast process that makes *in situ* studies extremely challenging. Noble solids krypton and xenon undergo a progressive pressure induced fcc to hcp martensitic transition with a very wide coexistence domain. Here, we took advantage of this unique feature to study the detailed mechanism of the transformation by employing *in situ* X-ray diffraction and absorption. We evidenced a four stages mechanism where the lattice mismatch between the fcc and hcp forms plays a key role in the generation of strain. We also determined precisely the effect of the transformation on the compression behavior of these materials.

# 1. Introduction

Martensitic phase transitions are fundamental first-order transformations that are, for instance, at the origin of steel hardening and shape memory effects [1]. They are diffusionless and proceed through the collective displacement of neighboring atoms. These subtle local atomic movements are at the origin of drastic changes in the materials' mechanical [2-4], transport [5] and electromagnetic [6] properties. Such transformations have been described in various materials such as metals [1,7] , semiconductors [8], Zr-based ceramics [9] and proteins [10] that are important for high-tech industry, planetary and life sciences. Despite significant research efforts have been invested in the past decades to explore the microscopic origin of martensitic transformations, the exact underlying mechanism is still debated. This lies in the very rapid nature of the first-order transformation that makes the monitoring of the phase transformation extremely challenging. For this reason, most of the experimental investigations were conducted *ex situ* using a variety of techniques such as optical microscopy, scanning electron microscopy, transmission electron microscopy (TEM) and X-ray diffraction [11-13], performed on (partially) transformed systems. However, a major obstacle lies in the capacity of such *ex situ* techniques to disentangle the different effects occurring during the fast phase transition when it proceeds. In particular, they do not give access to the variation of the atomic distances of the parent and martensitic phase during the transformation at the relevant conditions that may play a key role in the building-up and accumulation of strain in the material structure. Few *in situ* studies have been performed including neutron diffraction or TEM [14-17] but would require higher time resolution to capture the different steps of the transformation. Recently, the potential of proxy systems such as colloidal crystals to reveal kinetic features of martensitic transformations has been demonstrated [18].

The heavy noble solids xenon and krypton exhibit fundamental similarities when subjected to high pressures. Both systems undergo a pressure induced martensitic phase transition [19,20] with a wide coexistence domain between a face centered cubic (fcc) and an hexagonal close packed (hcp) structure. The existence of such a large coexistence domain (~60 GPa in xenon and ~400 GPa in krypton) is due to the proximity of the Gibbs free energies of the fcc and hcp forms. This is a fundamental feature among the martensitic transformations that provides a unique opportunity to follow the different stages of the transformation as it occurs. In addition, due to their simple electronic structure, noble solids are model systems to interpret such martensitic transitions at the microscopic level. Indeed, the fcc to hcp conversion in krypton and xenon which proceeds through one single shuffle step, is the easiest to model and has motivated several theoretical studies [21-23]. In xenon, a scenario was proposed in which the electronic $s$, $p$ to $d$ transitions leading to its metallization also induces the fcc to hcp conversion [24,25]. It was shown that the completion of the transformation occurs at 65 GPa in xenon and the extent of the coexistence domain is not affected by temperature annealing [19,20]. As recently reported in [26], a similar behaviour was observed in solid krypton but with a much wider pressure coexistence domain of ~400 GPa. Here again, the transition is not affected by temperature annealing attesting that it is not kinetically hindered. The extrapolated value of the complete fcc to hcp conversion is ~400 GPa [26].

The main objective of this experimental study is to exploit the progressive nature of the fcc to hcp conversion to *in situ* capture and characterize the different stages of the martensitic transformation by combining two complementary *in situ* probe techniques, X-ray diffraction and absorption that provide insights into the atomic arrangement at the bulk and local scale, respectively. We also aim at determining the effect of this transformation on the compression behavior of xenon and krypton.

## 2. Methods

All the experimental work was carried out at the X-ray absorption and X-ray diffraction beamlines ID24, BM23 and ID27 at the European Synchrotron Radiation Facility (ESRF) [27-29]. These three beamlines are optimal for performing very high-pressure experiments as they provide very intense, highly focused, high-energy X-ray beams. A list of the experimental conditions is presented in **Table I** of the main text.

### 2.1 XRD data acquisition and analysis

At the beamline ID27, monochromatic X-rays with wavelength λ=0.3738 Å were selected using a silicon (111) channel-cut monochromator and focused down to 3x3 μm$^2$ (FWHM) using a pair of Kirkpatrick-Baez (KB) mirrors [29]. The X-ray diffraction data were collected on a MAR165 planar CCD detector. A high purity cerium oxide powder was used as standard to accurately determine the integration parameters (detector tilt angles, sample to detector distance, detector beam center and instrumental peak broadening). Two kinds of diffraction images were acquired including continuous oscillation images over omega ($\Omega$) range of $\pm30°$ and step-oscillations images over the same $\Omega$ range using an $\Omega$ step size of 1°. The software CrysAlisPro was used to analyze the step-scan oscillation images [30]. The software Fit2D [31] was used to generate slices of 1-dimensional diffraction patterns along the azimuth for further Rietveld refinement of textured samples ($2\theta$ range up to 17°). The software MAUD [32] was employed to extract the unit cell volumes, volume fractions and coherently diffracting domain sizes (**Fig. S1** of the **Supplementary Material**).

Two independent XRD runs referred to as run-1 and run-2 were carried out at the beamline ID27. High purity xenon (99.999%) from the company Messer France was loaded in membrane diamond anvil cells (m-DACs) equipped with single-crystal diamonds of 150 µm and 300 µm culet size, respectively. A hole was drilled by laser machining on pre-indented rhenium gasket to serve as a pressure chamber. Prior to the xenon gas loading a ruby sphere (~3 µm in diameter), a 5 µm thick nickel foil and compressed gold powder pellet were placed in the pressure cavity. The pressure was measured using the standard ruby luminescence technique using the calibration of [33] and cross-checked using the unit-cell volume variation of nickel and gold using their well-established equation of states [33]. Nickel was only used as pressure calibrant at room temperature and prior to any laser annealing as it was shown that, in presence of noble gases, its compressional behaviour is substantially affected by high temperature treatment as noble gases can get incorporated in transition metals [26,34,35]. This material was also employed as YAG laser radiation absorber to evaluate the effect temperature annealing on the fcc-hcp transition in xenon. The annealing temperature was measured by analyzing the thermal emission as described in reference [36]. For clarity, considering the large amount of collected XRD data points ( > 100), we have reported the pressure-volume (*P-V*) data in **Table SI** of the **Supplementary Material**.

## 2.2 XAS data acquisition and analysis

The evolution of the local atomic arrangements of xenon and krypton with pressure were investigated at the two X-ray absorption (XAS) beamlines BM23 and ID24, respectively. BM23 is the only ESRF beamline that enables micro-XAS acquisitions at X-ray energies higher than 20 keV [28]. Therefore, this instrument is well suited for XAS studies in a DAC at the high energy K-edge of xenon (34.56 keV). It is equipped with a high-resolution Si (311) double-crystal monochromator and a Pt-coated KB-mirror system for X-ray beam focusing down to 5x5 $\mu m^2$ (FWHM) and higher harmonics rejection. Two ion chambers were used to measure the incident ($I_0$) and transmitted ($I_t$) intensities filled with appropriate gas mixtures of krypton and helium to achieve signal absorption of 30% and 70%, respectively. The local structure of krypton (K-edge energy of 14.385 keV) was investigated at the energy-dispersive XAS beamline ID24 [27]. At this instrument, a high intensity and highly focused X-ray beam of 10x10 $\mu m^2$ (FWHM) is generated by a Si(111) polychromator and rhodium-coated mirrors. The incident and transmitted intensities were recorded using a FReLoN CCD camera as described in Rosa et al. [26].

Three independent XAS runs referred to as run-3, run-4 and run-5 were performed. Run-3 was devoted to xenon and performed at BM23 while run-4 and -5 were focused on krypton and carried out at ID24 (examples of raw data are presented in **Fig. S2** and **Fig. S3** of the **Supplementary Material**). Typical data acquisition times were 30 minutes and 50 milliseconds at BM23 and ID24, respectively. For these runs all the m-DACs were equipped with nano-polycrystalline diamonds that enable acquiring glitch-free, high quality EXAFS data [37]. As for the experiments at ID27, a laser-drilled hole on a pre-indented rhenium gasket was employed as pressure chamber. The high purity noble gases were loaded together with a small ruby sphere used for pressure measurements. In run-3, at pressure higher than 90 GPa, the pressure was determined by XRD using the unit-cell volume variation of the rhenium gasket [38]. Diffraction data were acquired on two opposing positions to additionally evaluate the pressure gradient across the sample. For these measurements a MAR165 diffraction detector was installed at BM23 at a distance of 197.37 mm from the sample and XRD images were acquired without oscillation at a wavelength of 0.3594 Å (34.5 keV). The sample to detector distance, detector tilt and incident beam position were calibrated using a cerium oxide powder standard and the software Dioptas [39]. The dimensions of the diamonds culet size and the investigated pressure domains for the different runs are listed in **Table I**.

For both xenon and krypton, we used the same data reduction procedure to investigate the lattice distortion in the first and up to the third stacking layer. EXAFS spectra were normalized, converted from energy to $k$-space $\chi(k)$ and Fourier-transformed using a Kaiser-Bessel envelop from 3 to 10.5 Å$^{-1}$ and further analyzed without phase-shift analysis using the ARTHEMIS software package [40]. Each spectrum was either adjusted using fcc or hcp input models. Theoretical backscattering amplitude and phase shift functions for fcc and hcp phases were calculated for each pressure with the FEFF *ab initio* code included in ARTHEMIS. We used as an input model for the fitting procedure the crystallographic structures calculated from the equation of state of fcc and hcp Kr and Xe obtained from XRD. The scattering paths of the fcc and hcp forms in the first and second stacking layer exhibit an equivalent number of next-nearest neighbors and also bond distances if the two forms have the same unit cell volumes. The scattering paths employed in the fits covered also the third stacking layer in both forms. The latter is different in the two phases that exhibit either a ABCABC stacking sequence (fcc) or a ABABAB stacking sequence (hcp) (**Fig. 1, R5**).

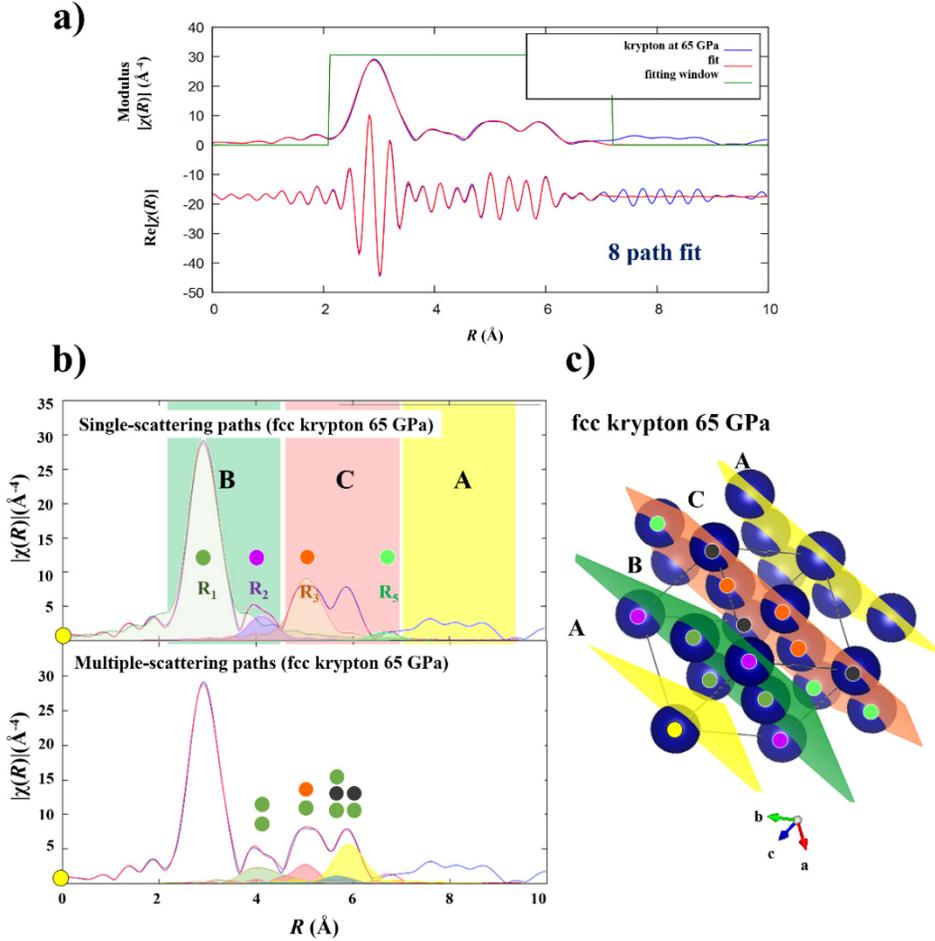

***Figure 1a) top panel:*** *Modulus of the krypton K-edge EXAFS function collected at 65 GPa representing the distribution of average inter-atomic distances (R) seen from a central absorbing atom. The modulus is obtained by Fourier transform of the normalized K-edge EXAFS functions (**Figures S2 and S3, Supplementary Material**) after their conversion from energy to k-space and multiplication by $k^2$ ($k^3(\chi(k))$). The amplitude of different peaks is proportional to the scattering probability. The blue lines represent the modulus (top) and the real part (bottom) of the original data, the green line represents the fitting window for the EXAFS analysis (phase shifted). The red lines represent the adjusted EXAFS spectrum.* ***b) bottom*** *panel: Modulus of the EXAFS function as above (blue line) plotted together with the moduli of the 4 most probable single-scattering paths (top) and the 4 most probable multiple-scattering paths (bottom) adjusted in the fit. **c)** For clarity, a section of a fcc structure with the stacking sequence ABCA is shown on the right. Atoms involved in the scattering paths emerging from the central atom (marked with a yellow circle) are highlighted with circles of different colour, corresponding to symbols and inter-atomic distances as marked in the EXAFS modulus plots in b) on the left.*

Krypton spectra were fitted only using the fcc input model, because the volume fraction of the hcp phase remains below 20% up to 65 GPa. The fits comprised 8 scattering paths similar to those employed in Filipponi and DiCicco [41], *i.e.,* four single scattering paths up to the 5th shell excluding the direct 4th shell scattering path due to its low probability (**Fig. 1**). Four multiple-scattering paths were included that have an important contribution to the EXAFS function such as the acute triangle in the first shell, the obtuse triangle including atoms in the third shell, and a forward and a double forward scattering path to atoms of the 4th shell. For each scattering path the inter-atomic distance ($R_i$) and its distribution ($\sigma^2_i$) were adjusted. A fitting window from 2 to 8 Å was chosen that resulted in 35 independent data points. The fit contained 17 adjusted parameters: 8 paths which included each individually fitted parameters $R_i$ and $\sigma^2_i$ and the energy shift $\Delta E$. The amplitude reduction factor was set to 1. The fits were performed with *k*-weighting of 1, 2, and 3 and background adjustments to diminish correlations between $\sigma^2$ and $R$, $\Delta E_0$. Fit results are listed in **Table SV**.

Xenon spectra were fitted up to 53 GPa with the fcc model and from 15 to 158 GPa with the hcp input structure. At low pressure up to 10 GPa, the fit window was set between 2 and 5.7 Å because of the significant damping of EXAFS oscillation amplitudes with *k* at these pressures. Fits contained 20 independent number of points and 13 adjusted parameters comprising 6 paths (4 single scattering similar to those for fcc Kr, two multiple scattering) with each individually fitted parameters $R$ and $\sigma^2$ and the energy shift $\Delta E$. Above this pressure the fits contained 60 independent points and 17 adjusted structural parameters comprising 8 paths with each individually fitted parameters $R$ and $\sigma^2$ and the energy shift $\Delta E$ and 16 fitted background parameters. The fit window was set between 2 and 8.5 Å. For all fits the amplitude reduction factor was set to 1. As for krypton all fits were performed with $k$-weighting of 1, 2, and 3 and background adjustments to diminish correlations between $\sigma^2$ and $R$, $\Delta E_0$. Fit results are listed in **Table SIV**.

# 3. Results and discussion

## 3.1 X-ray diffraction study of xenon

As presented in **Table I**, two independent X-ray diffraction (XRD) runs were carried out on solid xenon. Following the same methodology as for krypton [26], run-1 was devoted to the detailed investigation of the fcc to hcp martensitic transformation, the effect of temperature annealing and the determination of the equation of state (EoS) of fcc and hcp xenon, in a wide pressure regime up to 85 GPa. A large number of data points (75 data points) were collected in this run to increase the precision on the determination of the EoS parameters. As it was conducted in a large overlapping pressure interval (~ 40 GPa) with run-1, run-2 was used to control the consistency of the obtained results, increase their precision and robustness and further study the effect of temperature annealing on the fcc-hcp transition.

A series of diffraction images from run-1, recorded between 1 and 73.7 GPa, is presented in **Fig. 2.** At 1.09 GPa, a fcc single-crystal is observed. At this pressure, the intensity distribution in the (111) Bragg reflection is homogeneous and no trace of diffuse scattering is observed. At slightly higher pressure ($\Delta P \sim 0.3$ GPa), this diffraction peak largely deforms and a weak diffuse intensity is growing up in its vicinity. A careful analysis of the pressure induced intensity transfer from the fcc to the hcp phase along the [111] direction indicates that the appearance of X-ray diffuse signal is linked to the onset of the fcc/hcp martensitic transition. The presence of X-ray diffuse scattering was reported previously for xenon and krypton [19,20,26] and related to the presence of an increasing number of stacking faults (SFs) along the fcc [111] direction. For xenon, the onset of the fcc/hcp transition is precisely located at 4.9 GPa. As shown in **Fig. 2**, above this pressure, a progressive splitting of the (111) fcc reflection into the hcp (100), (002) and (101) reflections is recorded on panoramic XRD images. Single-crystal XRD measurements at this pressure confirmed that the martensitic phase transition in xenon produces an hcp phase with the expected orientation relation:  fcc (111) // hcp (0001) and fcc [1-11]// hcp [1-210].

Similarly to krypton [26] and as shown in **Fig. 2b)**, the X-ray diffuse signal present at ambient temperature remains very intense after temperature annealing (*T*>2400 K) in the entire pressure domain of fcc-hcp phase coexistence. This strongly suggests that the fcc-hcp transition and the presence hcp SFs are not due to non-homogeneous stress distribution in the pressure cavity. This is also supported by the deviatoric stress analysis of the XRD peak positions of the gold sample in run-1 and run-2 (**Fig. S4, S5, Supplementary Material**). This analysis reveals quasi-hydrostatic pressure conditions in the sample chamber up to ~15 GPa in run-1 and over the entire investigated pressure range in run-2. At pressures above ~15 GPa, a small deviatoric stress of 0.14 GPa develops in run-1 that increases to a moderate value of 0.5 GPa at 71 GPa. For comparison, this value is of the same order as in helium, a quasi-hydrostatic pressure medium [42], although an increase in strength with rare gas weight is expected. Martensitic transitions commonly induce shape changes of a transforming crystal during the collective shuffling of atoms [1]. This shape change can lead to the built up of internal stress in the crystal that is known as transformational stress [1]. The formation of multiple equivalent-sized martensite variants can counterbalance the shape change and limits thus the transformational stress [1,43]. The same mechanism could also release non-hydrostatic stress that is usually build in the diamond anvil cell by thinning of the gasket. The very low deviatoric stress observed in gold above the transition pressure suggests that this mechanism could be at work here.

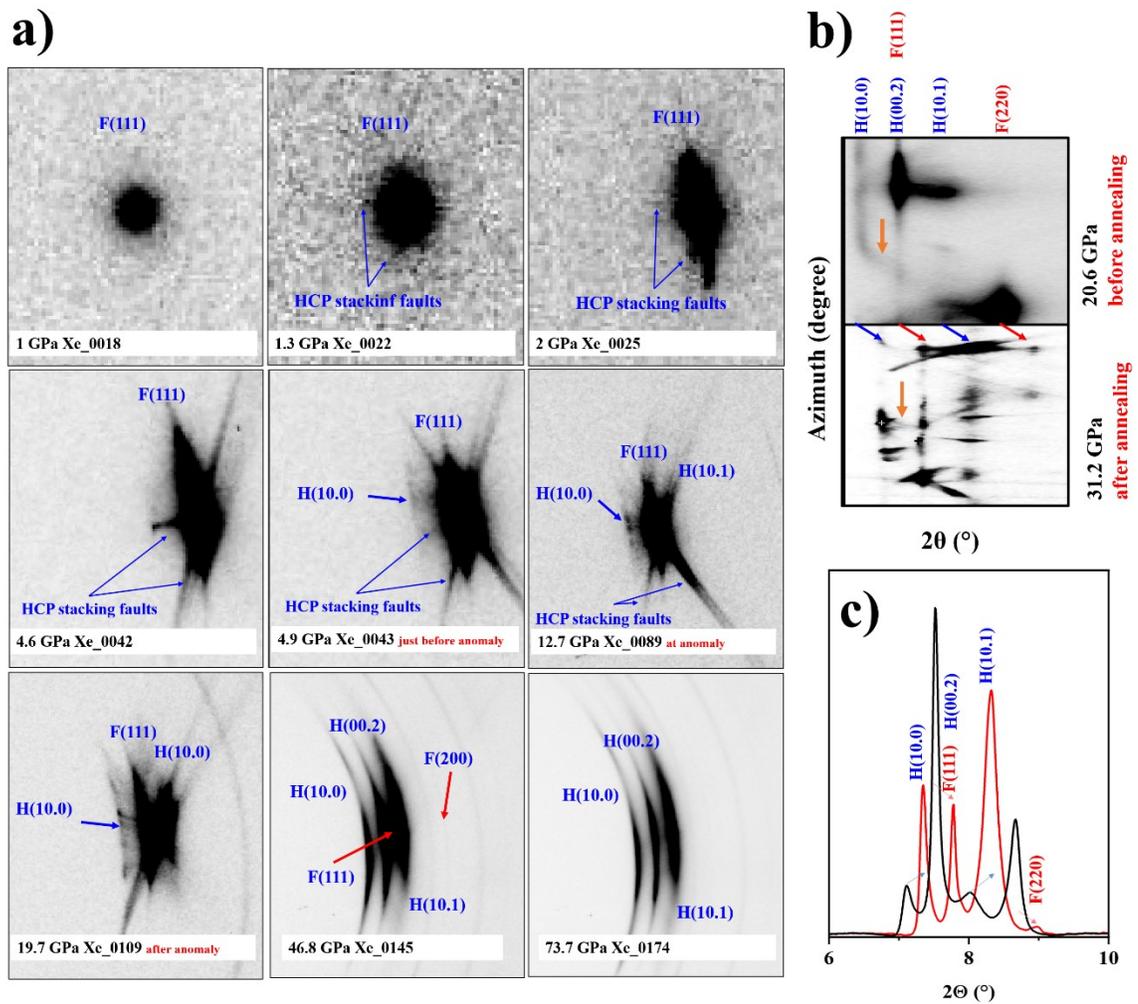

***Figure 2a) left panels:*** *Series of diffraction images from run-1 on pure xenon, showing the evolution of the xenon fcc (111) reflection (denoted as F(111)) with increasing pressure, the onset of the X-ray diffuse scattering linked to the emergence of the hcp phase and hcp Bragg reflections (denoted as H(10.0), H(00.2) and H(10.1)) after completion of the phase transition at ~74 GPa.* ***b)*** *and* ***c) right panels:*** *X-ray diffraction images (top) and corresponding integrated patterns (down) from run-2 on xenon before and after annealing at T = 2400 K. Note the pressure increase due to temperature annealing.*

## 3.2 Anomalous compression behavior

From run-1 and run-2, we have accurately determined the effect of the martensitic transformation on the equation of states (EoS) of fcc and hcp xenon. In particular, we examined the effect of the increasing concentration of stacking faults on these EoS. We have finely explored the low-pressure regime to better constrain the ambient pressure atomic volume ($V_0$) and examined the effect of temperature annealing. The pressure variation of the atomic volume of fcc and hcp xenon from this study are presented in **Fig. 3** and compared to the literature [20,44,45] in **Fig. S6 (Supplementary Material)**. The *P-V* data from run-1 and -2 are in excellent agreement in their overlapping pressure domain for both the fcc and hcp forms. As evidenced in **Fig. 3 and S6 (Supplementary Material)**, the present dataset exhibits lower dispersion than literature ones resulting in a more accurate determination of the equation of state parameters. From the absence of observable deviation of the atomic volume of xenon after high temperature treatment, we also could confirm that, similarly to krypton [19,26], temperature annealing has a negligible effect on the compression behavior of xenon.

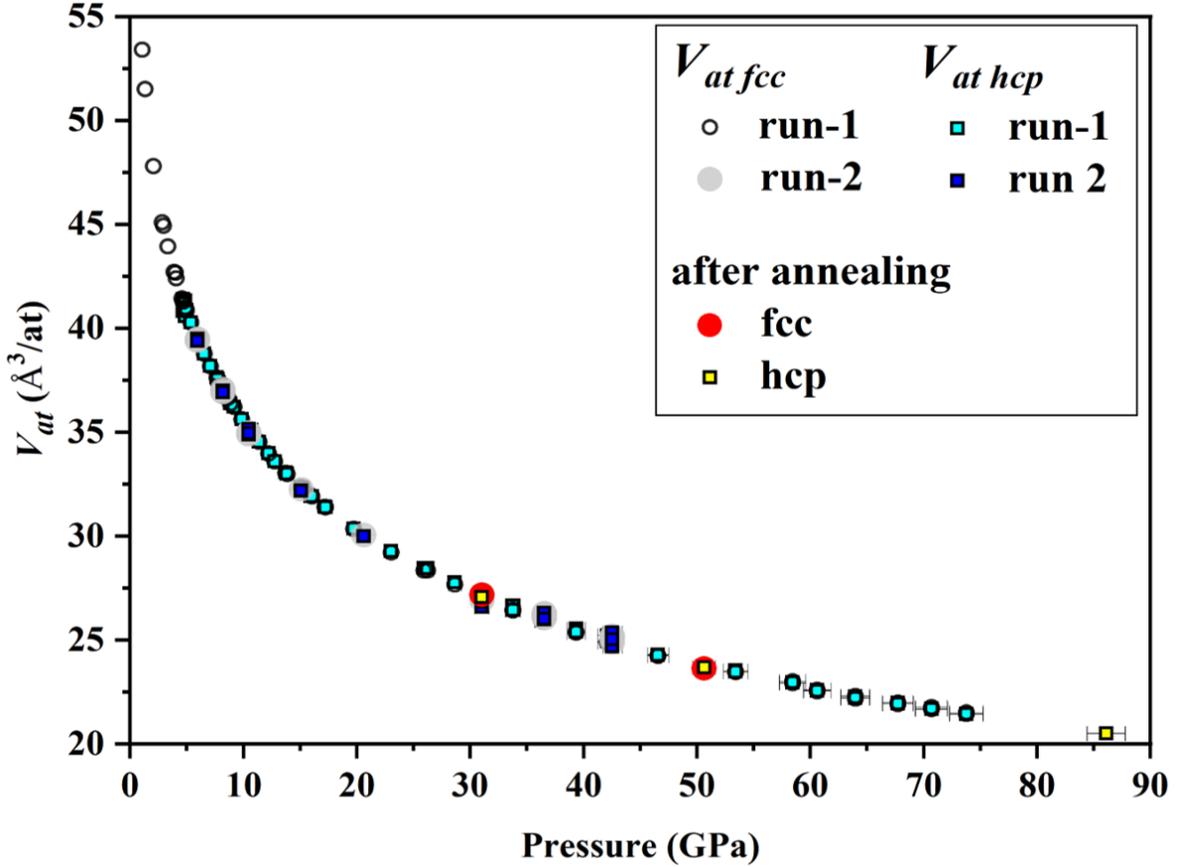

**Figure 3**. *Pressure evolution of the atomic volume $V_{at}$ ($Å^3$) for the fcc and hcp phases of xenon. The pressure has been determined using the ruby luminescence technique [33] up to 60 GPa beyond this pressure the pressure from the unit cell volume of gold was used and the EoS reported in [33]. Uncertainties on $V_{at}$ are in the order of $10^{-3}$ and smaller than the symbol size. The symbols for the fcc phase of run-2 and after annealing have been enlarged for better visualization but bear the same uncertainties as other points. Note that the effect of temperature annealing on $V_{at}$ is negligible and that there is little difference between $V_{at}$/fcc and $V_{at}$/hcp that is below 1% (**Table SI**). The literature data are presented in the left inset panel.*

The equation of state parameters, *i.e.* the unit-cell volume at ambient pressure $V_0$, bulk modulus $K_0$ and first derivative $K$' for fcc and hcp xenon were derived by adjusting the *P-V* data to a Vinet EoS using the software EoSFit [46], taking into account the uncertainties on $P$ and $V$. The obtained $V_0$, $K_0$ and $K'$ for fcc and hcp xenon are listed together with literature values in **Table SII** and **SIII (Supplementary Material),** respectively. They are in good agreement with the values of Dewaele et al. [47] but substantially deviate from the ones reported in reference [19,20]. This could originate from the high number of measured data points (75) in the present work, in particular at low pressure, where a slightly larger data dispersion can induce a large error in $V_0$. As for krypton [26], a near ideal $c/a$ ratio (1.63) is observed for all the hcp data points (**Table SI and Fig. S7, Supplementary Material**) attesting a similar reduction of the two crystallographic axis $a$ and $c$ of xenon with increasing pressure. This is indicative of a progressive martensitic transformation related to the increasing fraction and thickening of stacking faults in the material and rules out the existence of an intermediate orthorhombic close-packed structure as previously proposed[48,49].

Another source of uncertainty in the determination of the EoS parameters of xenon could arise from the nature of the fcc/hcp martensitic transformation. Indeed, it was shown that krypton exhibits an anomalous compression behavior related to the presence of an increasing number of hcp domains of nano-metric size [26]. This singular behavior was determined from the correlation between the pressure dependence of the volume fraction of the hcp phase and the anomaly in the normalized pressure, $F$, versus Eulerian strain, $f$ [50,51], where $f = [(V/V_0)^{-2/3}-1]/2$ and $F = P/(3f(1+2f)^{5/2})$. The $Ff$ plot analysis is a standard method to detect compression anomalies that are not obvious in the regular evolution of the compression curve ($P$-$V$ plot). We have conducted a similar $Ff$ plot analysis for xenon. As shown in **Fig. 4a)**, a clear deviation to a linear variation is evidenced in fcc xenon at ~15 GPa. As already mentioned, such a non-linear effect has also been determined in solid krypton at slightly higher pressure (~20 GPa)[26] suggesting a systematic evolution of the underlying microscopic mechanism during the fcc/hcp martensitic transition in heavy noble solids (**Fig. 4b)**).

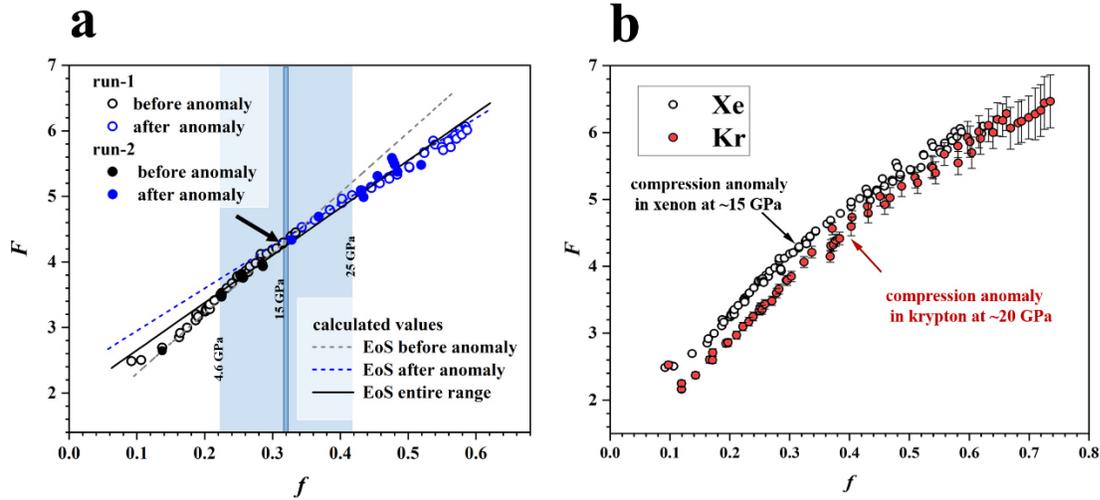

***Figure 4a)*** *Normalized pressure, F, vs. Eulerian strain f, for the fcc phase of xenon and **b)** compared to the one found for krypton in a previous work [26]. The grey and blue dashed lines correspond to calculated values from fitted EoS using the data obtained before (open symbols) and after (filled symbols) the anomaly. The black solid line corresponds to the data fitting over the entire P domain (up to 74 GPa). Uncertainties are 2% for the pressure and are within the symbol size for the Eulerian strain f and F. The interval where the compression anomaly occurs is highlighted by the light blue light shaded area, while the turn-over in compression behavior for xenon at ~15 GPa is indicated by a black arrow and dark blue line in **a)**. Note that we observed a similar behavior in krypton [26] as shown in **b)** where the turn-over in compression behavior is indicated by a red arrow.*

To correlate the observed $Ff$ plot anomaly with the hcp volume fraction ($VF_{hcp}$), we have determined its evolution with pressure using multi-phase Rietveld refinement of the XRD patterns using the intensity ratio of all Bragg reflections. As presented in **Fig. 5**, we found a progressive pressure variation of $VF_{hcp}$ reaching ~100 % at ~80 GPa. This behavior is contrasting with the one reported by Cynn *et al.* [20], Errandonea *et al.* [19] and Dewaele *et al.* [47] which exhibit a much sharper transformation. This is maybe related to the differences in analysis technique, *i.e.,* Rietveld refinement (this study) *versus* manual relative peak height analysis (previous works). It is worth noting that the complete fcc/hcp conversion in xenon occurs at much lower pressure than in krypton for which the pure hcp phase is expected at ~400 GPa. At 15 GPa, the pressure at which the anomaly in the *Ff* plot is observed, $VF_{hcp}$ reaches ~40 %, a value that is twice as large as for krypton ($VF_{hcp}$ at anomaly ~20 %). It is also worth noting that the volume fraction $VF_{hcp}$ is not significantly affected by temperature annealing.

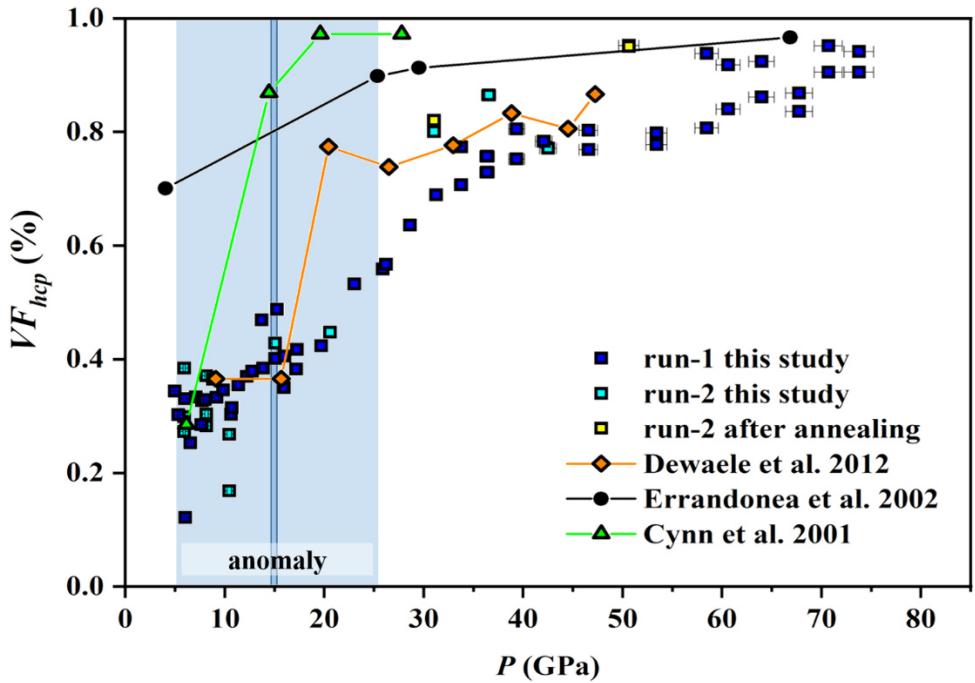

**Figure 5.** *Pressure evolution of the hcp xenon volume fraction ($VF_{hcp}$) extracted from Rietveld analysis of run-1 and run-2 (dark and light blue squares, respectively). Yellow filled squares indicate the data points obtained after temperature annealing in run-2. $VF_{hcp}$ from the literature obtained by Rietveld refinement of the intensity ratios of the fcc (200) and fcc (111) + hcp (002) reflections are indicated. Blue bar: pressure*

*at which the anomalous compression behavior is observed. At this pressure, $VF_{hcp}$ is ~40 %. In this study, the uncertainty on $VF_{hcp}$ is of the order of 6% and smaller than the symbol size.*

As presented in **Fig. 6**, we have also determined the pressure evolution of the coherently diffracting domain size for fcc ($CDS_{fcc}$) and hcp ($CDS_{hcp}$) xenon. These quantities were obtained by adjusting the XRD Bragg reflection profiles using Popa's analytical approximation implemented in the software MAUD [52]. We observe a strong reduction with pressure of $CDS_{fcc}$ while the $CDS_{hcp}$ is not significantly affected. This behavior is similar to the one previously observed for krypton [26]. This suggests a generic transformation mechanism in krypton and xenon at the first stages: the increasing fraction of nano-metric hcp SFs propagates, breaks fcc domains into smaller ones and interconnects gradually through the material at a regular pressure rate. The data suggest that the interconnection of hcp domains starting from ~5 GPa leads to the gradual deviation from a regular compression behavior in fcc xenon (**Fig. 4**). It is worth noting that this pressure coincides with the appearance of the first hcp diffraction peaks (**Fig. 2**). At pressures higher than 15 GPa beyond the inflexion point of the compression anomaly, the $CDS$ of the fcc phase does not reduce further contrary to its volume fraction, while the $CDS$ of the hcp phase increases slightly. These observations are similar to those made on krypton (**inset panel of Fig. 6**). This suggests that beyond the compression anomaly a modification in the fcc-hcp transformation mechanism occurs that balances the effect of transformational and compressional stresses.

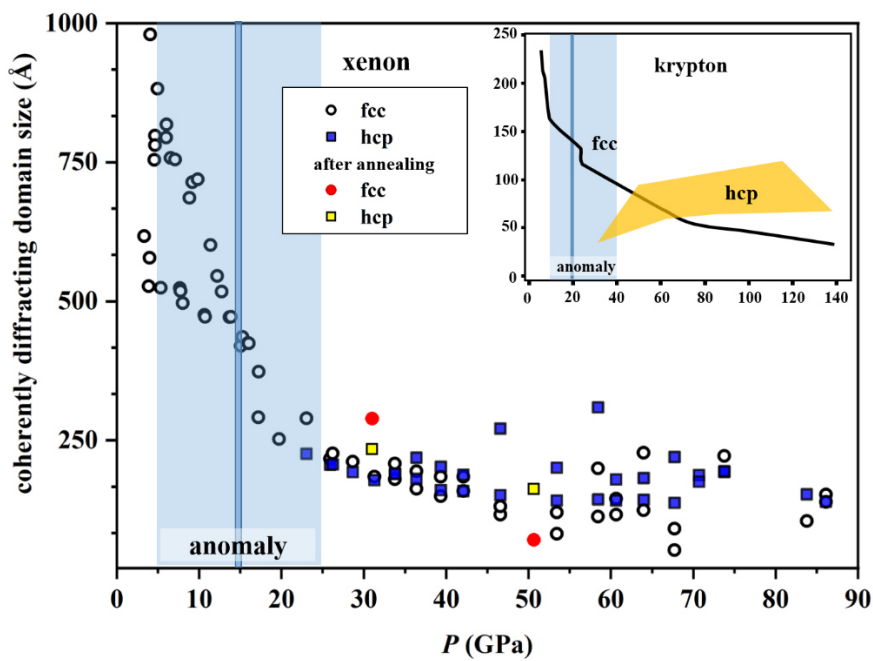

***Figure 6.*** *Pressure dependence of the coherently diffracting domain size (CDS) extracted by Rietveld analysis using Popa's analytical approximation for the fcc and hcp phases of xenon: circles and blue squares. The red and yellow symbols indicate the data points obtained after temperature annealing. The uncertainties on CDS extracted from Rietveld are of the order of 15 Å and smaller than the symbol size. The blue area highlights the pressure domain of the compression anomaly and the blue bar its inflection point. The pressure dependence of the CDS for the fcc and hcp krypton as reported in [26] is schematically represented in the inset panel.*

## 3.3 XAS study of krypton and xenon

In addition to XRD, we have carried out extended X-ray absorption fine structure (EXAFS) experiments to follow the pressure evolution of the local atomic arrangement of krypton and xenon during the martensitic transformation. EXAFS provides accurate structural data of the local atomic environment that complements the average information obtained by XRD. In the present study, noise-free EXAFS data of krypton and xenon were acquired in a diamond anvil cell (**Fig. S2 and S3, Supplementary Material**). Such measurements are now possible thanks to the recent development of μXAS facilities and nano-polycrystalline diamond anvil cells [27,28,37,53]. For both xenon and krypton, we used the same data analysis procedure to investigate inter-atomic distances $R_i$ between a central absorbing atom and neighboring atoms and their distributions $\sigma^2_i$. As represented in **Fig. 1**, the long-range EXAFS data were acquired up to a $k$ of 16 Å$^{-1}$ providing inter-atomic distances and their distributions up to the third stacking layer of the close-packed planes. As shown in **Fig. 1**, fcc and hcp structures respectively follow a ABCABCABC and ABABABAB pattern types. The scattering paths of the fcc and hcp forms in the first and second stacking layer exhibit an equivalent number of next-nearest neighbors and also bond distances if the two forms have the same unit cell volumes. Beyond the second stacking layer, several scattering paths for the two forms differ.

Long-range EXAFS is therefore well suited to probe the local structural modifications that occur during a martensitic transformation. Three EXAFS runs were performed. In run-3, we have collected high-quality EXAFS data of xenon in an extended pressure domain up to 155 GPa to monitor the structural deviations from the ideal fcc and hcp configurations during and after completion of the martensitic transformation. More data points were collected in run-5 and -6 on krypton to investigate with a higher resolution the martensitic transition in the vicinity of the compression anomaly (**Table I**). These data were also used to compare the inter-atomic potential forces to previously reported ones obtained using Monte-Carlo simulations [54].

For clarity, the resulting pressure variations of the extracted bond distances of the four scattering paths ($R_1$, $R_2$, $R_3$, $R_5$, **Fig. 1b**) and their distributions are reported in the **Table SIV and SV (Supplementary Material)** for xenon and krypton, respectively. The evolution of the normalized first nearest neighbor inter-atomic distance $R_1/R_0$ for fcc and hcp xenon obtained from the EXAFS and XRD analysis are presented in **Fig. 7a)**. Above ~25 GPa, the results of both approaches are in good agreement within their uncertainties. Below this pressure, the average first neighbor Xe-Xe inter-atomic distances extracted from the EXAFS data are slightly shorter than those obtained using XRD. However, no definitive conclusion could be drawn from the EXAFS data on xenon regarding the microscopic origin of the compression anomaly due to the few data points sampled in the pressure interval where it occurs (**Fig. 7a)**). In addition, in the low pressure regime, the inter-atomic interactions are dominated by van der Walls forces resulting in a reduction of the EXAFS oscillation amplitudes (**Fig. S2**) making their interpretation difficult. At pressures above 60 GPa, the first neighbor Xe-Xe distances extracted from EXAFS are slightly larger than those measured by XRD. Such positive deviations were previously related to the existence of vibrational modes in the direction normal to the Xe-Xe bond [55]. This suggests a modification of the local vibrational properties of xenon. The confirmation of this effect requires further analysis that goes beyond the scope of the present work.

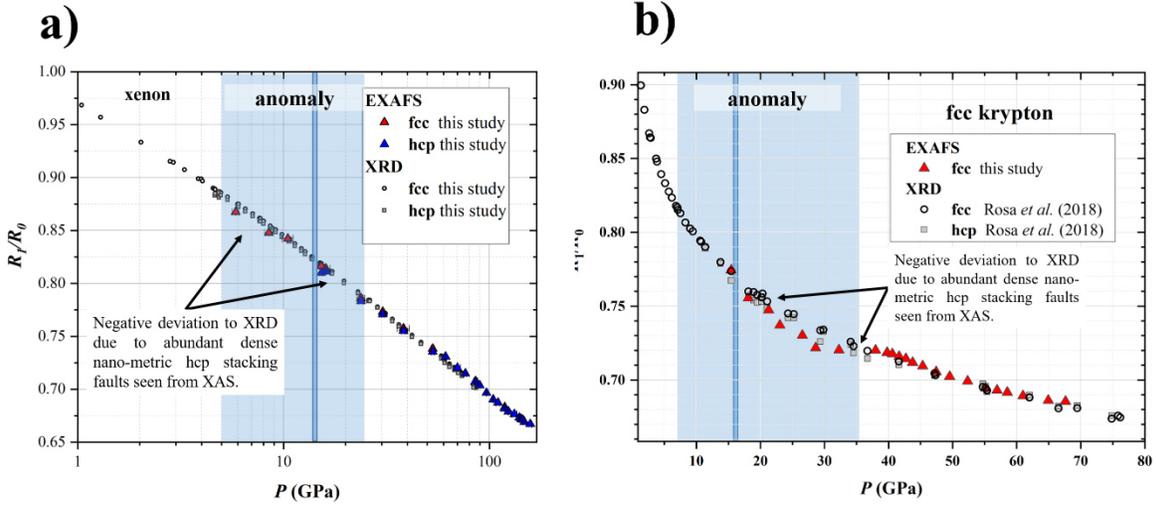

***Figure 7a)*** *Pressure evolution of the first nearest neighbor inter-atomic distances in fcc and hcp xenon normalized to its value at ambient pressure ($R_1/R_0$). Data obtained from EXAFS fitting using either the fcc or the hcp phase as input model are compared to calculated values using the EoS of fcc and hcp xenon derived from XRD in this study (Table II and III). $R_0$ was calculated from $V_0$ as obtained from the EoS of the present XRD data. A logarithmic pressure scale for xenon data was chosen to highlight the small deviations between results of different fit models and X-ray probes at low pressures. Uncertainties are smaller or in the order of the symbol sizes **b)** Pressure evolution of the first nearest neighbor inter-atomic distance of fcc krypton normalized to its value at ambient pressure ($R_1/R_0$). Data obtained from EXAFS analysis in this work are compared to those acquired from a previous XRD study [26]. $R_0$ values for fcc and hcp krypton were calculated from a previous EoS study [26]. The pressure domain of the compressional anomaly observed in this study in xenon and by Rosa et al. [26] in krypton are highlighted by a blue shaded area and the inflexion point of the anomaly is indicated with a blue bar.*

## 3.4 Origin of the compressional anomaly

The larger number of EXAFS data points collected on krypton in run-5 and 6 in the vicinity of the compression anomaly enabled a more detailed interpretation of its origin. As presented in **Fig. 7B**, from the EXAFS data, we observe a strong reduction of the first neighbor distance (here referred to as Kr-Kr) between ~18 and 36 GPa that is very difficult to assess from the XRD data [26]. This observation is related to the formation of a large amount of dense hcp stacking faults in fcc krypton and xenon in the pressure interval of the compression anomaly. These stacking faults exhibit a shorter next nearest neighbor distances than the parent fcc phase.  It is worth noting that the inflexion point of the compression anomaly in fcc krypton observed at ~20 GPa using XRD occurs at the onset of the negative deviation of Kr-Kr inter-atomic distances probed by EXAFS. At pressures below or above the compression anomaly, the Kr-Kr bond distances obtained using EXAFS and XRD (for the fcc phase, **Fig. 7B**) are in good agreement. The observations made on krypton are similar to those on xenon (**Fig. 7A**), but the deviations between inter-atomic distances extracted from EXAFS and XRD are more pronounced in krypton. This is explained by the smaller unit cell volume difference between the fcc and hcp forms in xenon than in krypton in the vicinity of the anomaly (see **Table SI, Supplementary Material**).

A similar conclusion can be drawn regarding the pressure evolution of the inter-atomic distances of the 2nd, 3rd and 5th next nearest neighboring atom extracted from EXAFS that deviates from the ideal fcc inter-atomic distances in the region of the compression anomaly for both elements (**Fig. S8**). Interestingly, far away from the anomaly at ~60 GPa, the inter-atomic distances in the second and third stacking layer ($R_2$, $R_3$ and $R_5$) are closer to those of the fcc than to those of the hcp phase obtained using XRD (**Fig. S8A**). This suggests the presence of a large fraction of boundary defects between hcp domains that have a local atomic arrangement close to a fcc lattice. Such defects could be located on hcp twins or grain boundaries.

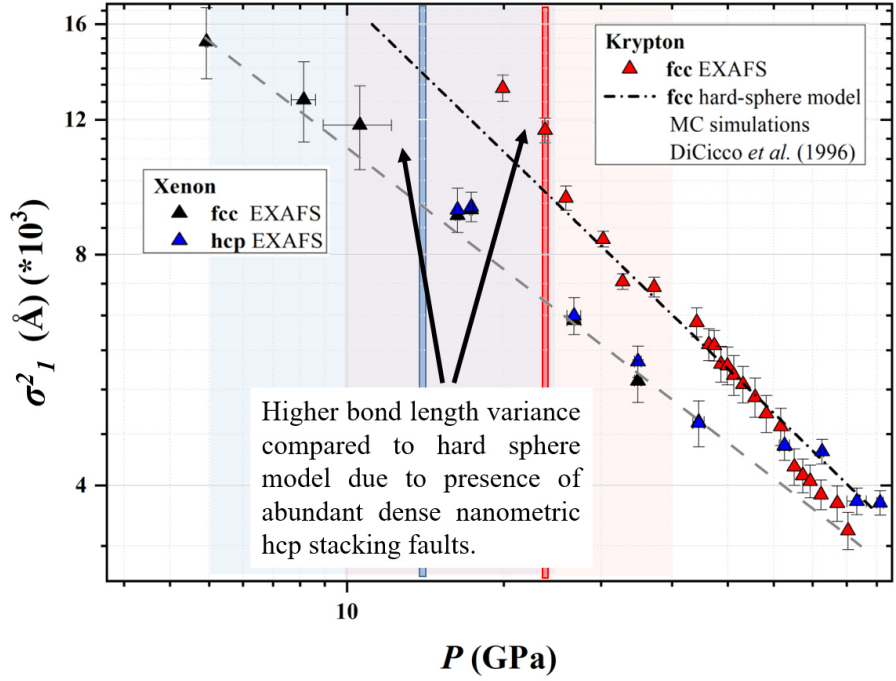

**Figure 8.** *Pressure variation of the fitted first neighbor distribution of krypton and xenon ($\sigma^2_1$) obtained from EXAFS. The data are compared to Monte Carlo simulation results using a hard-sphere model approach [54] for fcc krypton (denoted as MC in the figure legend). A linear trend is expected for a regular compression behavior. Due to the lack of computational studies for fcc/hcp xenon, we have drawn a linear line through the data points before and after the compression anomaly as a guide for the eyes.*

As shown in **Fig. 8**, the compression anomaly is also evidenced in the pressure variation of the first neighbor distance distribution $\sigma^2_1$ in krypton and to a lesser extent in xenon. In krypton, the fitted values of $\sigma^2_1$ up to 25 GPa show a strong deviation to the $\sigma^2_1$ calculated by Aziz and Slaman (HFD-B) and reported by DiCicco et al. [54] using a hard sphere model for fcc krypton. This deviation is reduced but still persistent up to ~ 40 GPa suggesting a high degree of structural disorder for fcc krypton in the pressure interval of the compression anomaly. A similar behavior occurs for xenon and for the 2$^{nd}$, 3$^{rd}$ and 5$^{th}$ nearest neighbors in krypton and xenon (**Fig. S9,**

**Supplementary Material**). Overall, the pressure dependence of $\sigma^2_1$ is consistent with the interpretation derived from the evolution of $R_i$: The increasing fraction of hcp SFs in the fcc structure induces a high degree of structural disorder. This effect is more pronounced in the pressure domain of the compression anomaly because the two structures have differing unit cell volumes at these conditions. Our conclusion is strengthened by our observations beyond the compression anomaly and up to 60 GPa: In this pressure domain we observe a reduction of $\sigma^2_1$ to the predicted trend of the atomic interaction potential in the hard sphere model of Aziz and Slaman [56]. Beyond 60 GPa, the bond distribution of krypton and xenon deviates from the linear variation expected for a regular compression behavior. In xenon, this behavior could be explained by the presence of a large fraction of boundary defects between hcp domains in the final sample micro-texture close to the completion of the transition.

## 4. Mechanism of the martensitic transformation

The combined results of EXAFS and XRD reveal a multi-stage growth of hcp SFs in the heavy noble gases krypton and xenon. The proposed microscopic mechanism of the martensitic transformation is presented in **Fig. 9.** Prior to the transformation, the starting material is a high quality single crystal with well-defined Bragg reflections and without detectable X-ray diffuse scattering that would be the signature of pre-existing defects (**Fig. 1**). In the first stage of the transition (stage 1 of **Fig. 9**), we observe a spontaneous formation of isolated but abundant hcp nano-metric stacking faults along the [111] fcc directions at relatively low pressures (~ 1.3 GPa and ~2.7 GPa for xenon and krypton, respectively). This is evidenced by the presence of diffused X-ray scattering signal in the vicinity of the fcc Bragg reflections (**Fig. 1**). This is conform to the general concept of the fcc-hcp martensitic transition developed by Olson and Cohen [57] proposing that the first step in martensitic nucleation is characterized by faulting on planes of closest packing resulting from the spontaneous formation of martensitic embryos. A mechanism responsible for the formation of these nano-metric SFs could be the recombination of two Shockley partial dislocations [1], which could generate two twin boundaries in the fcc lattice **(Fig. 9c)**. The formation of a large amount of SFs could explain the significant decrease of $CDS_{fcc}$ **(Fig. 9a)**.

The second stage of the transition is characterized by the interconnection and thickening of hcp SFs, evidenced by the appearance of distinct hcp diffraction peaks at approximately 5 GPa in xenon **(Fig. 1)**. Upon pressure increase, we observe a gradual formation of thin hcp domains that generate more intense XRD peaks **(Fig. 9b, stage 2)**. Single-crystal XRD indicates that all possible orientation of hcp domains, with (0001)hcp//(111)fcc (Shoji-Nishiyama orientation relations [1]) are observed; this represents four possible orientations, which we call variants here. As evidenced from EXAFS, in the pressure interval of the compression anomaly the hcp form exhibits shorter next-nearest neighbour distances (up to 4% difference for krypton) than the parent fcc phase leading to significant local disorder and strain in the structure **(Fig. 7, 8)**. It is likely that the presence and interconnection of these hcp domains with smaller unit cell volume induces a strain increase in the bulk material that is, in turn, responsible for the observed compression anomaly. This feature is of particular importance as it might be similar to the one responsible for the hardening of metals (i.e. steel hardening) **(Fig. 3**, **Fig. 9A, stage 2)**. We emphasize that this critical observation can only be made *in situ* and not on quenched materials as the unit cell volumes and local structures are substantially modified during the quenching process. It is also worth noting that the compression behavior of the gold sample located in the pressure cavity is regular over the entire pressure domain (**Fig. S4, S5, Supplementary Material**). This provides additional evidence that the observed compression anomaly in krypton and xenon is intrinsically due to the fcc-hcp lattice mismatch and not to the presence of pressure gradients in the sample chamber.

The third stage of the transformation, after the compression anomaly, is more specific to krypton and xenon. This might be because in steel hardening processes, the quenching of the material would be performed at the stage 2 of the transformation, when the concentration of strain is maximum. At this stage, the fcc-hcp lattice difference vanishes and the fcc phase retrieves a regular compression behavior (**Fig. 3, 7** and **8**). This is also a clear indication that the hcp-fcc lattice mismatch is at the origin of the strain enhancement and compression anomaly. This stage is characterized by a stagnation of the $CDS_{fcc}$ and $CDS_{hcp}$ and an increase of the hcp volume fraction $VF_{hcp}$ (**Fig. 9a**). The fourth stage which is characterized by a constant evolution of the $CDS_{fcc}$ and $VF_{hcp}$ at high pressure might be due to the collective shuffling of remaining fcc domains surrounded by hcp SFs. It was shown [1,43] that the formation of equivalent-sized hcp variants in metals could compensate for the transformational and the compressive strains induced by the martensitic transition. This could produce a selection of variants to minimize strain (transformational and non-hydrostatic stress naturally present in the DAC), which could not detect here due to limited access of reciprocal space with XRD in the DAC. Such a mechanism would lead to a large amount of domain boundaries in the resulting micro-texture (**Fig. 9b), stage 4**). This is consistent with the EXAFS results on xenon beyond 60 GPa that show the prevalence of a "fcc-like" local atomic arrangement (**Fig. 7, 8 and S8, S9 of the Supplementary Material**). As represented in **Fig. 9c),** the hcp domain boundaries with inverted stacking sequence indeed exhibit "fcc-like" interfaces. Both krypton and xenon preserve the initial strong preferred orientation of the hcp phase (**Fig. 1**) up to the maximum investigated pressure. This suggests that the relative lattice rotations of isolated hcp grains induced by slip systems other than those along the (111) crystallographic planes play a minor role in the transformation.

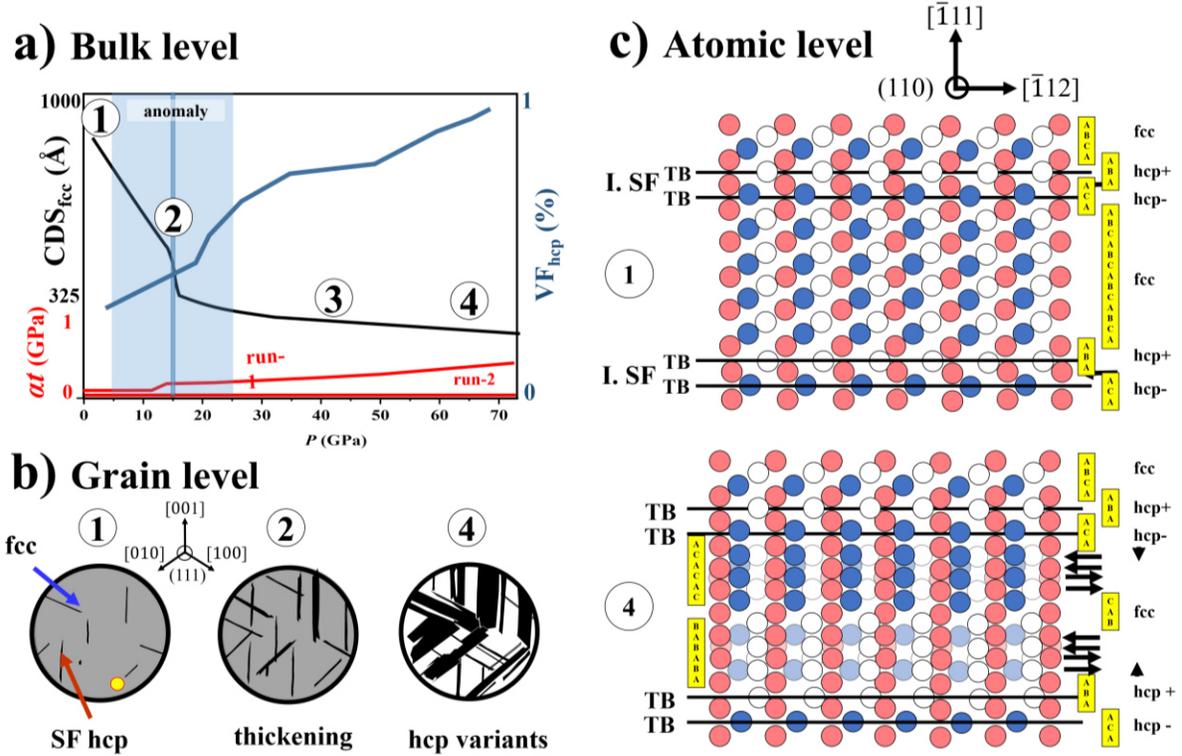

**Fig. 9.** *Evolution of mechanical and microstructural parameters during the martensitic fcc/hcp transition in xenon at* **a)** *the bulk,* **b)** *the grain and* **c)** *the atomic level. The transition is divided in 4 transformational stages: 1) nano-metric Hcp stacking fault (SF) formation from 1.3 GPa onward in xenon, 2) interconnection of SF that induces the compressional anomaly, 3) transformation of remaining fcc domains via shuffling into hcp variant domains of equivalent size and 4) final microstructure of coexisting hcp variants with inverted stacking sequences. In* **a)** *the pressure dependence of CDS*fcc*, stress in gold (αt) and VF*hcp *are summarized. The pressure domain of the compression anomaly is highlighted by a blue shaded area and the inflection point of the anomalous compression behavior is delineated by a blue bar.* **b)** *summarizes the progression of the transition at the grain scale for the different stages. The yellow circle represents the size*

*of the X-ray beam. At stage 4, hcp variant domains are shown in black and white, to highlight differences in the hcp stacking sequence of either ABABA or ACACA (see **c**) for more details). **c**) Stacking-fault formation and thickening mechanisms at the atomic scale that can explain the observations from both XRD and EXAFS. Note that the stacking sequence of close-packed planes along the [111] is shown in upward direction. Abbreviations are: TB: twin boundary, I. SF: inverse stacking fault, hcp+ (ABA, stacking sequence) and hcp- (ACACA, stacking sequence).*

# 5. Conclusion

We have examined in details the fcc to hcp martensitic transformation in solid krypton and xenon using *in situ* XRD and EXAFS. We have evidenced a multi-stage mechanism in which the transition initially proceeds through the spontaneous nucleation of hcp nanometric domains. This stage is followed by a strain accumulation in the bulk material due to fcc-hcp lattice mismatch and the presence of a large fraction of hcp stacking faults. In this transient domain, an anomaly is evidenced in the equation of state of both materials. In the following stage of the transformation, the lattice mismatch vanishes and a normal compression behavior is retrieved. The micro-texture of xenon in the final stage is characterized by the presence of coexisting hcp variants with inverted stacking sequences. The possible formation of equi-sized hcp variants could explain the low deviatoric stress as seen by the gold pressure marker. The present work reveals the multiple stages of the transformation and disentangles the different effects occurring during the transition. We also provide evidence that the variation of the atomic distances of the parent and martensitic phase during the transformation plays a key role in the building-up and accumulation of strain in the materials' structure.

**Acknowledgements**

We would like to acknowledge the ESRF for providing beamtime for this work and we are grateful for his help of J. Jacobs during the gas-loading of the samples. We also thank F. Perrin and S. Pasternak for their technical assistance during the experiments. Lawrence Livermore National Laboratory is operated by Lawrence Livermore National Security, LLC, for the U.S. Department of Energy, National Nuclear Security Administration under contract No. DE-AC52-07NA27344.

## Author Contributions

The original idea was conceived by A.D.R. The experiments were performed by A.D.R., G.G., V.S., G.M., M.K., R.B. and O.M.. T.I. provided the nano-polycrystalline diamonds for the X-ray absorption experiments. The data were analysed by A.D.R., A.D., G.G., V.S., F.De A. and O.M.. The manuscript was written by A.D.R and M.A.B with contributions from all the co-authors.

## Competing Interests statement

The authors declare no competing interests.

## Figure Legends and Tables.

**Figure 1. a) top panel:** Modulus of the krypton $K$-edge EXAFS function collected at 65 GPa representing the distribution of average inter-atomic distances ($R$) seen from a central absorbing atom. The modulus is obtained by Fourier transform of the normalized $K$-edge EXAFS functions (**Figures S2 and S3, Supplementary Material**) after their conversion from energy to $k$-space and multiplication by $k^2$ ($k^3(\chi(k))$). The amplitude of different peaks is proportional to the scattering probability. The blue lines represent the modulus (top) and the real part (bottom) of the original data, the green line represents the fitting window for the EXAFS analysis (phase shifted). The red lines represent the adjusted EXAFS spectrum. **b) bottom panel**: Modulus of the EXAFS function as above (blue line) plotted together with the moduli of the 4 most probable single-scattering paths (top) and the 4 most probable multiple-scattering paths (bottom) adjusted in the fit. **c)** For clarity, a section of a fcc structure with the stacking sequence ABCA is shown on the right. Atoms involved in the scattering paths emerging from the central atom (marked with a yellow circle) are highlighted with circles of different colour, corresponding to symbols and inter-atomic distances as marked in the EXAFS modulus plots in b) on the left.

**Figure 2. a) left panels:** Series of diffraction images from run-1 on pure xenon, showing the evolution of the xenon fcc (111) reflection (denoted as F(111)) with increasing pressure, the onset of the X-ray diffuse scattering linked to the emergence of the hcp phase and hcp Bragg reflections (denoted as H(10.0), H(00.2) and H(10.1)) after completion of the phase transition at ~74 GPa. **b)** and **c) right panels:** X-ray diffraction images (top) and corresponding integrated patterns (down) from run-2 on xenon before and after annealing at $T$ = 2400 K. Note the pressure increase due to temperature annealing.

**Figure 3.** Pressure evolution of the atomic volume $V_{at}$ (Å$^3$) for the fcc and hcp phase of xenon. The pressure has been determined using the ruby luminescence technique [33] up to 60 GPa beyond this pressure the pressure from the unit cell volume of gold was used and the EoS reported in [33]. Uncertainties on $V_{at}$ are in the order of $10^{-3}$ and smaller than the symbol size. The symbols for the fcc phase of run-2 and after annealing have been enlarged for better visualization but bear the same uncertainties as other points. Note that the effect of temperature annealing on $V_{at}$ is negligible and that there is little difference between $V_{at}/fcc$ and $V_{at}/hcp$ that is below 1% (**Table SI**). The literature data are presented in the left inset panel.

**Figure 4.** Normalized pressure, $F$, vs. Eulerian strain $f$, for the fcc phase of xenon **a)** compared to the one found for krypton in a previous work [26] **b)**. The grey and blue dashed lines correspond to calculated values from fitted EoS using the data obtained before (open symbols) and after (filled symbols) the anomaly. The black solid line corresponds to the data fitting over the entire P domain (up to 74 GPa). Uncertainties are 2% for the pressure and are within the symbol size for the Eulerian strain $f$ and $F$. The interval where the compression anomaly occurs is highlighted by the light blue light shaded area, while the turn-over in compression behavior for xenon at ~15 GPa is indicated by a black arrow and dark blue line in **a)**. Note that we observed a similar behavior in krypton [26] as shown in **b)** where the turn-over in compression behavior is indicated by a red arrow.

**Figure 5.** Pressure evolution of the hcp xenon volume fraction ($VF_{hcp}$) extracted from Rietveld analysis of run-1 and run-2 (dark and light blue squares, respectively). Yellow filled squares indicate the data points obtained after temperature annealing in run-2. $VF_{hcp}$ from the literature obtained by Rietveld refinement of the intensity ratios of the fcc (200) and fcc (111) + hcp (002) reflections are indicated. Blue bar: pressure at which the anomalous compression behavior is observed. At this pressure, $VF_{hcp}$ is ~40 %. In this study, the uncertainty on $VF_{hcp}$ is of the order of 6% and smaller than the symbol size.

**Figure 6.** Pressure dependence of the coherently diffracting domain size ($CDS$) extracted by Rietveld analysis using Popa's analytical approximation for the fcc and hcp phases of xenon: circles and blue squares. The red and yellow symbols indicate the data points obtained after temperature annealing. The uncertainties on $CDS$ extracted from Rietveld are of the order of 15 Å and smaller than the symbol size. The blue area highlights the pressure domain of the compression anomaly and the blue bar its inflection point. The pressure dependence of the CDS for the fcc and hcp krypton as reported in [26] is schematically represented in the inset panel.

**Figure 7a)** Pressure evolution of the first nearest neighbor inter-atomic distances in fcc and hcp xenon normalized to its value at ambient pressure ($R_1/R_0$). Data obtained from EXAFS fitting using either the fcc or the hcp phase as input model are compared to calculated values using the EoS of fcc and hcp xenon derived from XRD in this study (Table II and III). $R_0$ was calculated from $V_0$ as obtained from the EoS of the present XRD data. The logarithmic pressure scale for xenon data was chosen to highlight the small deviations between results of different fit models and X-ray probes at low pressures. Uncertainties are smaller or in the order of the symbol sizes. **b)** Pressure evolution of the first nearest neighbor inter-atomic distance of fcc krypton normalized to its value at ambient pressure ($R_1/R_0$). Data obtained from EXAFS analysis in this work are compared to those acquired from a previous XRD study [26]. $R_0$ values for fcc and hcp krypton were calculated from a previous EoS study [26]. The pressure domain of the compressional

anomaly observed in this study in xenon and by Rosa et al. [26] in krypton are highlighted by a blue shaded area and the inflexion point of the anomaly is indicated with a blue bar.

**Figure 8.** Pressure variation of the fitted first neighbor distribution of krypton and xenon ($\sigma^2{}_1$) obtained from EXAFS. The data are compared to Monte Carlo simulation results using a hard-sphere model approach [54] for fcc krypton (denoted as MC in the figure legend). A linear trend is expected for a regular compression behavior. Due to the lack of computational studies for fcc/hcp xenon, we have drawn a linear line through the data points before and after the compression anomaly as a guide for the eyes.

**Fig. 9.** Evolution of mechanical and microstructural parameters during the martensitic fcc/hcp transition in xenon at **a)** the bulk, **b)** the grain and **c)** the atomic level. The transition is divided in 4 transformational stages: 1) nano-metric Hcp stacking fault (SF) formation from 1.3 GPa onward in xenon, 2) interconnection of SF that induces the compressional anomaly, 3) transformation of remaining fcc domains via shuffling into hcp variant domains of equivalent size and 4) final microstructure of coexisting hcp variants with inverted stacking sequences. In **a)** the pressure dependence of $CDS_{fcc}$, stress in gold ($\alpha t$) and $VF_{hcp}$ are summarized. The pressure domain of the compression anomaly is highlighted by a blue shaded area and the inflection point of the anomalous compression behavior is delineated by a blue bar. **b)** summarizes the progression of the transition at the grain scale for the different stages. The yellow circle represents the size of the X-ray beam. At stage 4, hcp variant domains are shown in black and white, to highlight differences in the hcp stacking sequence of either ABABA or ACACA (see **c)** for more details). **c)** Stacking-fault formation and thickening mechanisms at the atomic scale that can explain the observations from both XRD and EXAFS. Note that the stacking sequence of close-packed planes along the [111] is shown in upward direction. Abbreviations are: TB: twin boundary, I. SF: inverse stacking fault, hcp+ (ABA, stacking sequence) and hcp- (ACACA, stacking sequence).

**Table I.** List of the experimental conditions. BL: beamline, SC: single crystal, NPD: nano-polycrystalline diamond. Raw data are presented in **Fig. S1-3**.

| Element | run number | BL | technique | beam size ($\mu m^2$) | beam energy (keV) | pressure range (GPa) | pressure standard | diamond type |
|---------|-----------|------|-----------|-----------------------|-------------------|----------------------|-------------------|--------------|
| Xenon | 1 | ID27 | XRD | 3x3 | 33.17 | 0-86 | Au, Ni, | SC |
|       | 2 |      |           |       |       |       | ruby | SC |
|       | 3 | BM23 | EXAFS | 5x5 | 34.56 | 0-155 | ruby, Re | NPD |
| Krypton | 5 | ID24 | EXAFS | 10x10 | 14.33 | 15-68 | ruby | NPD |
|       | 6 |      |           |       |       |       |      |     |

# Supplementary material

# X-ray study of krypton and xenon under pressure reveals the mechanism of martensitic transformations

A.D. Rosa[1*], A. Dewaele[2,3], G. Garbarino[1], V. Svitlyk[1], G. Morard[4], F. De Angelis[5], M. Krstulovic[6], R. Briggs[7], T. Irifune[8], O. Mathon[1] and M.A. Bouhifd[9]

1. *European Synchrotron Radiation Facility (ESRF), 71, Avenue des Martyrs, Grenoble, France.*

2. *CEA, DAM, DIF, 91297 Arpajon Cedex, France*

3. *Université Paris-Saclay, CEA, Laboratoire Matière en Conditions Extrêmes, 91680 Bruyères-le-Châtel, France*

4. *Université Grenoble Alpes, Université Savoie Mont Blanc, CNRS, IRD, IFSTTAR, ISTerre, 38000 Grenoble, France*

5. *Dipartimento di Fisica, Universita' di Roma La Sapienza - Piazzale Aldo Moro 5, 00185 Roma, Italy*

6. *University of Potsdam, Institute of Geosciences, Karl-Liebknecht-Str. 24-25, 14476 Potsdam-Golm, Germany*

7. *Lawrence Livermore National Laboratory, Livermore, CA, United States of America*

8. *Geodynamics Research Center, Ehime University, 2-5 Bunkyo-cho, Matsuyama 790-8577, Japan*

9. *Laboratoire Magmas et Volcans, Université Clermont Auvergne, CNRS, IRD, OPGC, F-63000 Clermont-Ferrand, France*

*corresponding author: angelika.rosa@esrf.fr

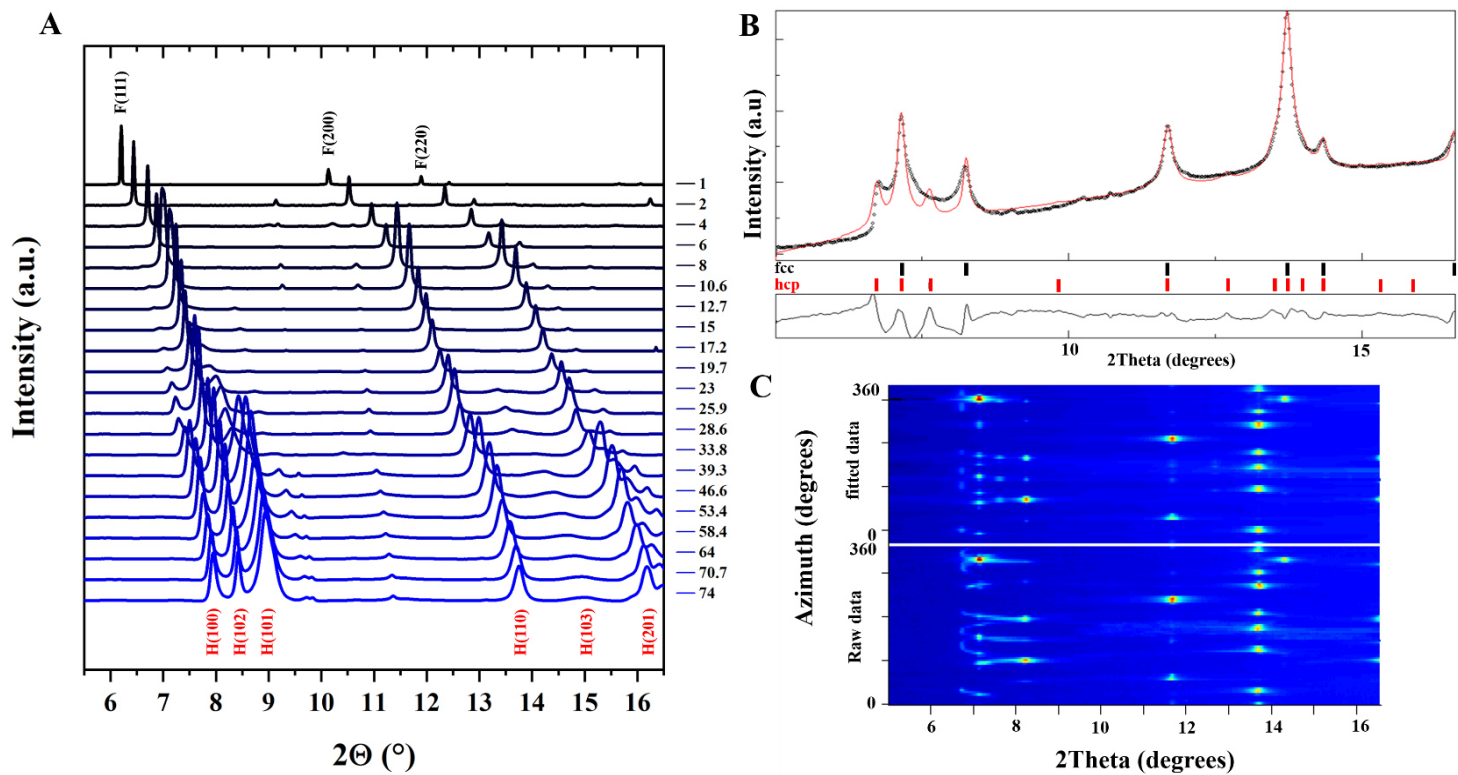

**Figure S1. A** Series of integrated diffraction patterns with most intense peaks labelled from 1 to 74 GPa. The labels F and H indicate the diffraction peaks from fcc and hcp xenon, respectively. **B** Rietveld refinement of a typical diffraction pattern collected at 10.6 GPa. **C.** Azimuthally enrolled image showing the sample texture at 10.6 GPa (images were sliced and integrated along the azimuth into 72 spectra every 5°).

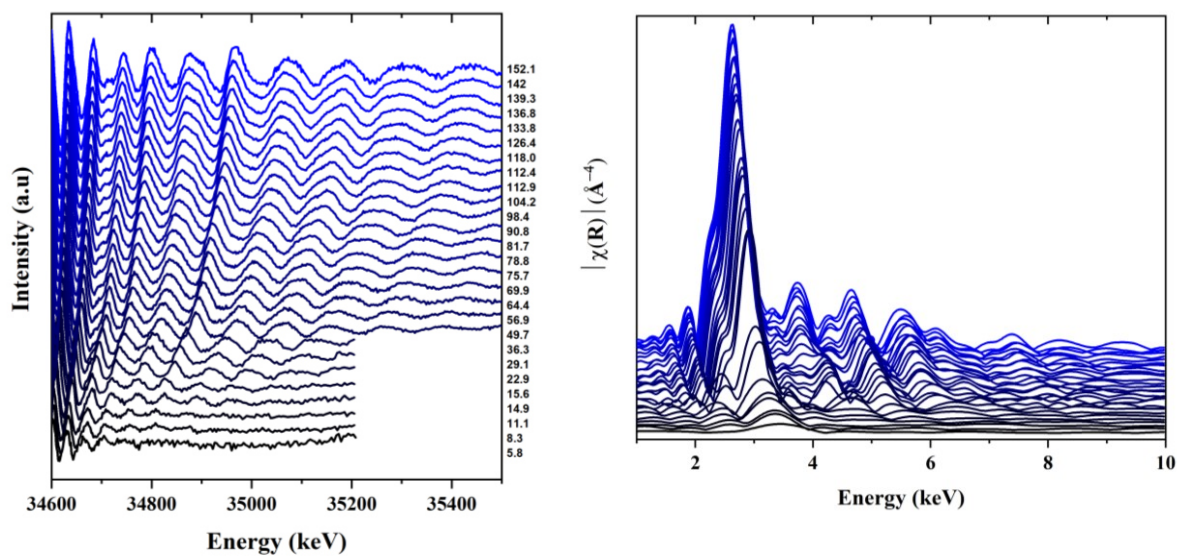

**Figure S2. Left and right panels:** Evolution of normalized Xe K-edge EXAFS spectra and their Fourier transformed with pressure.

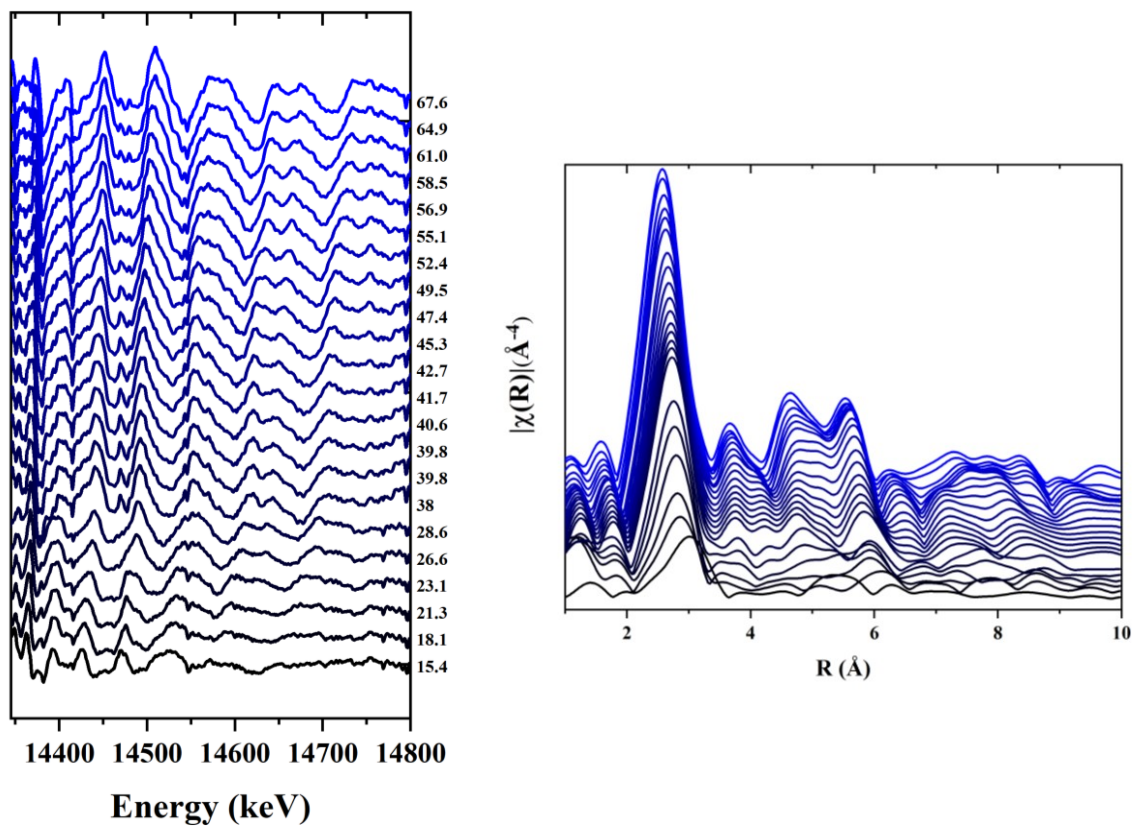

**Figure S3. Left and right panels:** Evolution of normalized Kr K-edge EXAFS spectra and their Fourier transformed with pressure.

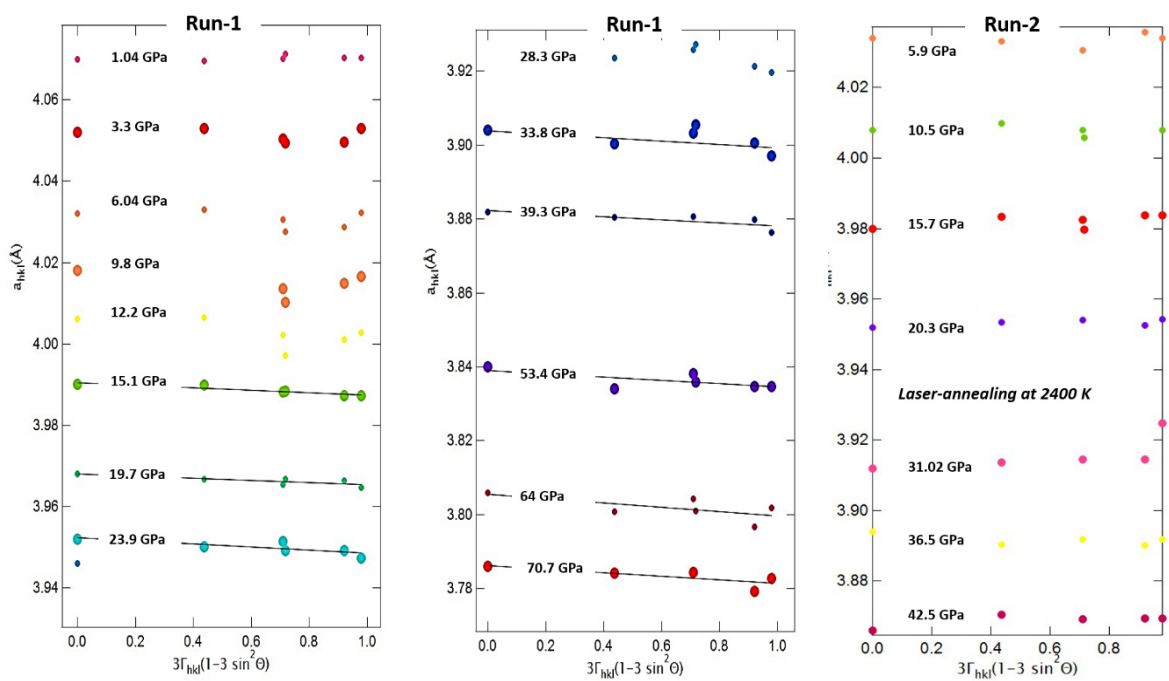

**Figure S4.** Pressure dependence of the lattice parameter $a$ of gold loaded together with xenon in run-1 and run-2 derived for observed reflections (hkl) and plotted against $3\Gamma_{hkl}$ (1-3sin$^2\theta$), where $\theta$ is the 2theta position of the reflection and $\Gamma_{hkl} = (h^2k^2 + k^2l^2 + l^2h^2)/(h^2 + k^2 + l^2)^2$. The indicated pressure is the one derived from ruby. The plots are used to determine the stress ($t$) evolution in gold with pressure similar to [1]. Under hydrostatic conditions, the lattice parameter determined from different reflections is equivalent for gold. This situation is observed in run-1 until 13.9 GPa and run-2 throughout the entire pressure range. The measured lattice parameter systematically deviates from the hydrostatic value in presence of non-hydrostatic stress depending on the *hkl* indices. This situation is observed in run-1 beyond 13.9 GPa. The value of $t$ was determined from the slope $M_1$ and intercept $M_0$ and presented in Fig. S6.

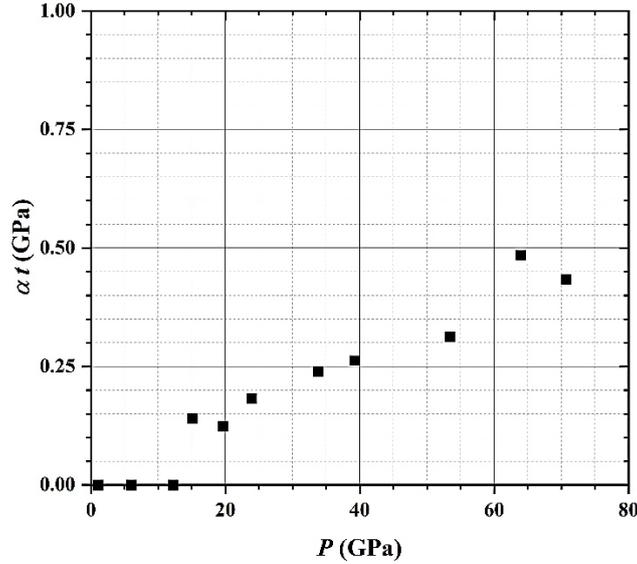

**Fig. S5**. Deviatoric stress ($t$) as a function of pressure in run-1 derived from the analysis of $a_{hkl}$ versus its $2\theta$ position in gold (**Fig. S4**) and the relation $t \sim -3M_1/(\alpha M_0 S)$, where $M_1$ and $M_0$ are determined from the slope of the $a_{hkl}$ *versus* $3\Gamma_{hkl}$ (1-3sin$^2\theta$) variation (**Fig. S4**). In the present analysis, $\alpha$ was assumed to be 0.5 corresponding to uniaxial compression conditions expected in the DAC experiments. $S$, the elastic anisotropy factor, was taken from [2]. For more details, see [1].

**Table SI.** Pressure evolution of the unit cell parameters $a$ and $c$, $c/a$ ratio, and volume fraction of the hcp phase ($VF_{hcp}$) extracted from Rietveld refinement of two independent experimental runs (run-1 and run-2). The uncertainties on pressure, evaluated from the pressure differences before and after an acquisition, measured with the ruby luminescence gauge [3], is smaller than 0.5% up to 100 GPa. The intrinsic uncertainties of the ruby, Au and Ni metal pressure gauges are 2%. The relative uncertainty on atomic unit cell volume is smaller than $10^{-3}$ over the entire pressure range. The error on the extracted volume fraction for the hcp phase ($VF_{hcp}$) are in the order of 6%. The error on the c/a ratio is in the order below $10^{-5}$ for the regular XRD data and below $10^{-3}$ for the XRD maps.

| Xenon | $P_{ruby}$ (GPa) | $a_{Ni}$ (Å) | $a_{Au}$ (Å) | $P_{Au/Ni}$ (GPa) | $a_{Xe\ FCC}$ (Å) | $a_{Xe\ HCP}$ (Å) | $c_{Xe\ HCP}$ (Å) | $c/a$ ratio | $VF_{hcp}$ (%) |
|---|---|---|---|---|---|---|---|---|---|
| **run 1** | | | | | | | | | |
| 1 | 1.04(1) | 3.5172(1) | 4.0699(1) | 1.09(1) | 5.9776(1) | | | | |
| 2 | 1.29(3) | 3.5140(1) | 4.0672(1) | 1.46(2) | 5.9061(1) | | | | |
| 3 | 2.03(4) | 3.5091(1) | 4.0600(1) | 2.40(2) | 5.7610(1) | | | | |
| 4 | 2.81(7) | 3.5036(1) | 4.0536(1) | 3.31(5) | 5.6502(1) | | | | |
| 5 | 2.91(4) | | | | 5.6432(1) | | | | |

| # | | | | | | | | | |
|---|---|---|---|---|---|---|---|---|---|
| 6 | 3.30(4) | 3.5004(1) | 4.0498(1) | 3.84(6) | 5.6013(1) | | | | |
| 7 | 3.85(8) | 3.4993(1) | 4.0461(1) | 4.2(1) | 5.5487(1) | | | | |
| 8 | 3.98(5) | | | | 5.5470(1) | | | | |
| 9 | 4.04(1) | | | | 5.5355(1) | | | | |
| 10 | 4.55(9) | | | | 5.4929(1) | | | | |
| 11 | 4.64(1) | 3.4922(1) | 4.0416(1) | 5.2(2) | 5.4883(1) | 3.8660(1) | 6.3072(2) | 1.631 | |
| 12 | 4.65(2) | | | | 5.4860(1) | 3.8747(1) | 6.3687(2) | 1.644 | |
| 13 | 4.86(6) | | | | 5.4741(1) | 3.8827(1) | 6.3345(2) | 1.631 | |
| 14 (map 1) | 4.86(6) | | | | 5.4716(1) | 3.8767(1) | 6.3301(2) | 1.633 | |
| 15 | 4.86(6) | | | | 5.4704(1) | 3.8574(1) | 6.3189(2) | 1.638 | |
| 16 | 4.86(6) | | | | 5.4672(1) | 3.8583(1) | 6.2972(2) | 1.632 | |
| 17 | 4.96(4) | | | | 5.4678(1) | 3.8671(1) | 6.3189(2) | 1.632 | 0.34 |
| 18 | 5.34(3) | 3.4883(1) | 4.0363(1) | 5.9(2) | 5.4416(1) | 3.8482(1) | 6.2972(2) | 1.632 | 0.30 |
| 19 | 5.98(5) | 3.4846(1) | 4.0318(1) | 6.6(2) | 5.4005(1) | 3.8199(1) | 6.2339(2) | 1.632 | 0.33 |
| 20 | 6.04(3) | | | | 5.4000(1) | 3.8195(1) | 6.2332(2) | 1.632 | 0.12 |
| 21 | 6.53(2) | 3.4844(1) | 4.0287(1) | 6.82(5) | 5.3726(1) | 3.8001(1) | 6.2006(2) | 1.631 | 0.25 |
| 22 | 7.07(4) | | | | 5.3446(1) | 3.7808(1) | 6.1688(2) | 1.632 | 0.33 |
| 23 | 7.64(5) | 3.4781(1) | 4.0256(1) | 7.7(3) | 5.3189(1) | 3.7614(1) | 6.1416(2) | 1.633 | 0.28 |
| 24 | 7.72(3) | 3.4761(1) | 4.0225(1) | 8.1(3) | 5.3164(1) | 3.7592(1) | 6.1376(2) | 1.633 | 0.33 |
| 25 | 7.99(10) | | | | 5.3016(1) | 3.7493(1) | 6.1178(2) | 1.632 | 0.33 |
| 26 | 8.62(4) | 3.4718(1) | 4.0173(1) | 8.9(3) | 5.2732(1) | 3.7306(1) | 6.0853(2) | 1.631 | |
| 27 | 8.82(8) | | | | 5.2596(1) | 3.7205(1) | 6.0701(2) | 1.631 | 0.36 |
| 28 | 9.15(5) | 3.4700(1) | 4.0142(1) | 9.4(2) | 5.2517(1) | 3.7150(1) | 6.0612(2) | 1.632 | 0.33 |
| 29 | 9.83(6) | 3.4669(1) | 4.0110(1) | 9.9(2) | 5.2229(1) | 3.6949(1) | 6.0274(2) | 1.631 | 0.35 |
| 30 | 10.62(6) | 3.4640(1) | 4.0080(1) | 10.5(3) | 5.1976(1) | 3.6765(1) | 5.9977(2) | 1.631 | 0.30 |
| 31 | 10.71(5) | | | | 5.1977(1) | 3.6765(1) | 5.9983(2) | 1.632 | 0.31 |
| 32 | 11.36(7) | 3.4604(1) | 4.0035(1) | 11.2(3) | 5.1693(1) | 3.6570(1) | 5.9647(2) | 1.631 | 0.35 |
| 33 | 12.20(3) | 3.4568(1) | 3.9995(1) | 11.9(3) | 5.1412(1) | 3.6373(2) | 5.9321(1) | 1.631 | 0.37 |
| 34 | 12.74(6) | 3.4599(1) | 3.9975(1) | 11.8(2) | 5.1222(1) | 3.6239(1) | 5.9198(1) | 1.631 | 0.38 |
| 35 | 13.7(1) | 3.4507(1) | 3.9924(1) | 13.2(3) | 5.0936(1) | 3.6043(1) | 5.8789(1) | 1.631 | 0.47 |
| 36 | 13.83(2) | | | | 5.0912(1) | 3.6022(1) | 5.8735(1) | 1.631 | 0.38 |
| 37 | 15.1(2) | 3.4421(1) | 3.9858(1) | 14.8(7) | 5.0583(1) | 3.5791(1) | 5.8327(1) | 1.630 | 0.40 |
| 38 | 15.24(2) | | | | 5.0546(1) | 3.5764(1) | 5.8080(2) | 1.630 | 0.49 |

Change in compression behaviour

| # | | | | | | | | | |
|---|---|---|---|---|---|---|---|---|---|
| 39 | 15.9(1) | 3.4418(1) | 3.9818(2) | 15.2(3) | 5.0365(1) | 3.5627(2) | 5.8078(1) | 1.630 | 0.35 |
| 40 | 16.02(1) | | | | 5.0351(1) | 3.5627(1) | 5.8080(1) | 1.630 | 0.41 |
| 41 | 17.2(1) | 3.4383(1) | 3.9767(2) | 16.0(2) | 5.0083(1) | 3.5444(1) | 5.7775(2) | 1.630 | 0.38 |
| 42 | 17.23(3) | | | | 5.0074(1) | 3.5431(1) | 5.7754(1) | 1.630 | 0.42 |
| 43 | 19.70(1) | 3.4302(1) | 3.9649(1) | 18.12(1) | 4.9512(1) | 3.5040(1) | 5.7123(1) | 1.630 | 0.42 |
| 44 | 23.0(1) | 3.4184(1) | 3.9483(1) | 21.3(3) | 4.8890(1) | 3.4618(1) | 5.6423(1) | 1.630 | 0.53 |
| 45 | 25.9(2) | 3.4081(1) | 3.9333(1) | 24.2(7) | 4.8404(1) | 3.4292(1) | 5.5885(1) | 1.630 | 0.56 |
| 46 | 26.22(4) | | | | 4.8404(1) | 3.4294(1) | 5.5882(1) | 1.630 | 0.57 |
| 47 | 28.63(19) | 3.3973(1) | 3.9225(1) | 26.9(5) | 4.8012(1) | 3.4022(1) | 5.5434(2) | 1.629 | 0.64 |
| 48 | 31.26(5) | 3.3868(1) | 3.9105(1) | 29.8(6) | 4.7637(1) | 3.3752(2) | 5.5010(2) | 1.630 | 0.69 |
| 49 | 33.76(7) | 3.3773(1) | 3.9002(1) | 32.5(5) | 4.7297(1) | 3.3532(2) | 5.4743(1) | 1.633 | 0.77 |
| 50 | 33.84(7) | | | | 4.7279(1) | 3.3458(1) | 5.4551(1) | 1.630 | 0.71 |

| # | map | | | | | | | | | |
|---|---|---|---|---|---|---|---|---|---|---|
| 51 | | 36.37(6) | 3.3688(1) | 3.8902(1) | 35.1(6) | 4.6991(1) | 3.3296(2) | 5.4309(1) | 1.631 | 0.76 |
| 52 | | 36.43(6) | | | | 4.6972(1) | 3.3243(1) | 5.4207(1) | 1.631 | 0.73 |
| 53 | | 39.3(1) | 3.3584(1) | 3.8789(1) | 38.2(6) | 4.6658(1) | 3.3052(1) | 5.3989(1) | 1.633 | 0.81 |
| 54 | | 39.4(1) | | | | 4.6637(1) | 3.3017(1) | 5.3821(1) | 1.630 | 0.75 |
| 55 | | 42.1(1) | 3.3489(1) | 3.8682(1) | 41.3(6) | 4.6355(1) | 3.2855(1) | 5.3957(1) | 1.642 | |
| 56 | | 42.2(1) | | | | 4.6354(1) | 3.2807(2) | 5.3491(1) | 1.630 | 0.78 |
| 57 | | 46.6(2) | 3.3366(1) | 3.8544(1) | 45.5(6) | 4.5938(1) | 3.2549(2) | 5.2965(2) | 1.627 | 0.80 |
| 58 | | 46.8(2) | | | | 4.5948(1) | 3.2520(1) | 5.3019(1) | 1.630 | 0.77 |
| 59 | | 53.4(1) | 3.3183(1) | 3.8348(1) | 51.9(6) | 4.5452(1) | 3.2190(1) | 5.2443(1) | 1.629 | 0.78 |
| 60 | | 53.5(1) | | | | 4.5443(1) | 3.2161(1) | 5.2436(1) | 1.630 | 0.80 |
| 61 | | 58.44(7) | 3.3066(2) | 3.8213(1) | 56.4(7) | 4.5149(1) | 3.1954(1) | 5.1893(1) | 1.624 | 0.94 |
| 62 | | 58.50(6) | | | | 4.5098(1) | 3.1919(1) | 5.2052(2) | 1.631 | 0.81 |
| 63 | | | 3.2964(1) | 3.8092(1) | 60.6(9) | 4.4879(1) | 3.1757(1) | 5.1733(1) | 1.629 | 0.92 |
| 64 | | | | | 60.6(9) | 4.4835(1) | 3.1730(2) | 5.1739(1) | 1.631 | 0.84 |
| 65 | | | 3.2886(1) | 3.7999(1) | 64.0(9) | 4.4684(1) | 3.1616(1) | 5.1496(1) | 1.629 | 0.92 |
| 66 | | | | | 64.0(9) | 4.4601(1) | 3.1566(1) | 5.1485(2) | 1.631 | 0.86 |
| 67 | | | 3.2799(1) | 3.7900(1) | 67.7(1.2) | 4.4472(1) | 3.1471(1) | 5.1243(1) | 1.628 | 0.87 |
| 68 | | | | | 67.7(1.2) | 4.4424(1) | 3.1441(1) | 5.1253(1) | 1.630 | 0.84 |
| 69 | | | 3.2735(2) | 3.7822(1) | 70.7(1.4) | 4.4324(1) | 3.1351(1) | 5.1045(1) | 1.628 | 0.95 |
| 70 | | | | | 70.7(1.4) | 4.4262(1) | 3.1319(1) | 5.1021(1) | 1.629 | 0.91 |
| 71 | | | 3.2668(2) | 3.7745(1) | 73.8(1.5) | 4.4150(1) | 3.1222(1) | 5.0829(2) | 1.628 | 0.91 |
| 72 | | | | | 73.8(1.5) | 4.4104(1) | 3.1198(1) | 5.0836(1) | 1.629 | 0.94 |

Annealed on one side at 1800 K

| # | map | | | | | | | | | |
|---|---|---|---|---|---|---|---|---|---|---|
| 73 | | | 3.2374(2) | | 86.1(1.7) | | 3.0705(1) | 5.0103(2) | 1.632 | 1.00 |
| 74 | map 2 | | 3.2375(2) | | 86.1(1.7) | | 3.0731(1) | 5.0194(1) | 1.633 | 1.00 |
| 75 | | | 3.2422(2) | | 86.1(1.7) | | 3.0720(1) | 5.0169(2) | 1.633 | |

run2

| # | map | | | | | | | | | |
|---|---|---|---|---|---|---|---|---|---|---|
| 76 | | 5.92(1) | 3.4882(1) | 4.0315(1) | 6.2(2) | 5.4059(1) | 3.822(3) | 6.245(1) | 1.634 | 0.38 |
| 77 | map 3 | 5.92(1) | | | | 5.4044(1) | 3.821(3) | 6.238(1) | 1.633 | 0.27 |
| 78 | | 5.92(1) | | | | 5.4013(1) | 3.820(3) | 6.235(2) | 1.632 | 0.30 |
| 79 | | 8.16(5) | 3.4746(1) | 4.0199(1) | 8.4(2) | 5.2933(2) | 3.739(2) | 6.113(2) | 1.635 | 0.37 |
| 80 | map 4 | 8.16(5) | | | | 5.2882(2) | 3.738(2) | 6.108(1) | 1.634 | 0.30 |
| 81 | | 8.16(5) | | | | 5.2857(2) | 3.738(2) | 6.102(1) | 1.633 | 0.28 |
| 82 | | 10.5(1) | 3.4672(1) | 4.0067(1) | 10.3(2) | 5.1916(2) | 3.680(1) | 5.995(1) | 1.629 | |
| 83 | map 5 | 10.5(1) | | | | 5.1872(2) | 3.668(1) | 5.989(1) | 1.633 | 0.17 |
| 84 | | 10.5(1) | | | | 5.1902(2) | 3.671(1) | 5.9903(8) | 1.632 | 0.27 |
| 85 | | 15.07(3) | 3.4463(1) | 3.9831(1) | 14.6(1) | 5.0530(3) | 3.5726(8) | 5.8239(3) | 1.630 | 0.43 |
| 86 | | 20.6(1) | 3.4224(1) | 3.9556(1) | 20.0(1) | 4.9350(3) | 3.4885(7) | 5.6933(3) | 1.632 | 0.45 |
| 87 | | 31.0(4) | 3.3846(2) | 3.9117(1) | 30.0(3) | 4.7720(3) | 3.3720(4) | 5.4986(2) | 1.631 | 0.82 |
| 88 | | 31.0(4) | | | | 4.7572(3) | 3.3552(6) | 5.5012(2) | 1.640 | |
| 89 | map 6 | 31.0(4) | | | | 4.7743(3) | 3.3581(5) | 5.4931(2) | 1.636 | 0.54 |
| 90 | | 31.0(4) | | | | 4.7663(4) | 3.3404(3) | 5.5035(2) | 1.648 | 0.86 |
| 91 | | 31.0(4) | | | | 4.7680(3) | 3.3626(1) | 5.4795(2) | 1.630 | 0.97 |

Annealing on one side at 2400 K

| # | map | | | | | | | | | |
|---|---|---|---|---|---|---|---|---|---|---|
| 92 | map 7 | 36.5(5) | 3.3719(2) | 3.8912(1) | 34.5(9) | 4.7122(1) | 3.3288(1) | 5.4346(5) | 1.633 | 0.96 |

| | | | | | | | | |
|---|---|---|---|---|---|---|---|---|
| 93 | 36.5(5) | | | 4.7108(1) | 3.3304(1) | 5.4231(4) | 1.628 | 0.92 |
| 94 | 36.5(5) | | | 4.7167(1) | 3.3312(1) | 5.4286(4) | 1.630 | 0.86 |
| 95 | 36.5(5) | | | 4.7203(1) | 3.3392(1) | 5.4501(5) | 1.632 | 0.67 |
| 96 | 36.5(5) | | | 4.7056(1) | 3.3263(1) | 5.4283(5) | 1.632 | |
| 97 | 36.5(5) | | | 4.7095(1) | 3.3289(1) | 5.4224(5) | 1.629 | 0.86 |
| 98 | 42.5(1) | 3.3521(1) | 39.7(5) | 4.6496(1) | 3.2837(2) | 5.3599(3) | 1.632 | 0.76 |
| 99 | 42.5(1) | | | 4.6446(1) | 3.2848(1) | 5.3579(3) | 1.631 | 0.80 |
| 100 map8 | 42.5(1) | | | 4.6501(3) | 3.2902(1) | 5.3699(3) | 1.632 | 0.71 |
| 101 | 42.5(1) | | | 4.6538(3) | 3.2986(1) | 5.3826(3) | 1.632 | |
| 102 | 42.5(1) | | | 4.6342(3) | 3.2715(1) | 5.3285(4) | 1.629 | |
| 103 | 42.5(1) | | | 4.6444(3) | 3.2850(1) | 5.3584(2) | 1.631 | 0.81 |
| annealing on one side at 2100 K | | | | | | | | |
| 104 | 50.6(1) | | | 4.5549(4) | 3.2229(1) | 5.2653(2) | 1.634 | 0.95 |

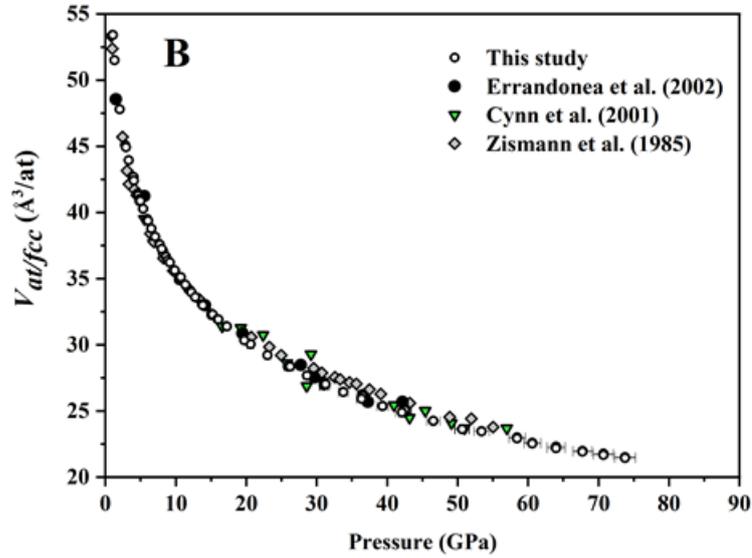

**Fig.S6.** Comparison of the pressure evolution of the atomic volume $V_{at}/fcc$ (Å³) from this study and the literature [4-6].

**Table SII.** Equation of state for fcc xenon: model and fit parameters $V_0$, $K_0$, $K'$, pressure range and number of data points from this work and the literature.

| fcc xenon | model | $V_0$ (Å³) | $K_0$ (GPa) | $K'$ | $P$ range (GPa) |
|---|---|---|---|---|---|
| **This study** | Vinet | 235.15(9) | 4.54(2) | 6.26(1) | 1 – 74 GPa (75 points) |
| **Dewaele et al. 2012 [7]** | R.-Vinet[1] (0 K) | 233.76 | 4.887 | 6.18(5) | 4 – 50 (20 points) |
| **Errandonea et al. 2002 [4]** | BM[2] 3rd | 249.75 | 4.3(6) | 5.7(5) | 1.5 – 41 (8 points) |
| **Cynn et al. 2001 [5]** | BM[2] 3rd | 252.21 | 3.6(5) | 5.5(4) | 3-50 GPa (29 points) |

[1]R.-Vinet - Rose–Vinet
[2]BM - Birch-Murnaghan

**Table SIII.** Equation of state for hcp xenon: model and fit parameters $V_0$, $K_0$, $K'$, the pressure range, number of data points from this work and the literature.

| Hcp xenon | model | $V_0$ (Å³) | $K_0$ (GPa) | $K'$ | $P$ range (GPa) |
|---|---|---|---|---|---|
| **This study** | Vinet | 118.71 | 4.24(4) | 6.35(3) | 4.6 - 86 GPa (65 points) |
| **Dewaele et al. 2012 [7]** | R.-Vinet[1] (0 K) | 114.82 | 4.887 fixed | 6.2955 | 10 - 260 GPa |
| **Cynn et al. 2001 [5]** | BM[2] 3rd | 126.10 | 4.3(3) | 4.9(1) | 52-127 GPa (15 points) |

[1]R.-Vinet - Rose–Vinet
[2]BM - Birch-Murnaghan

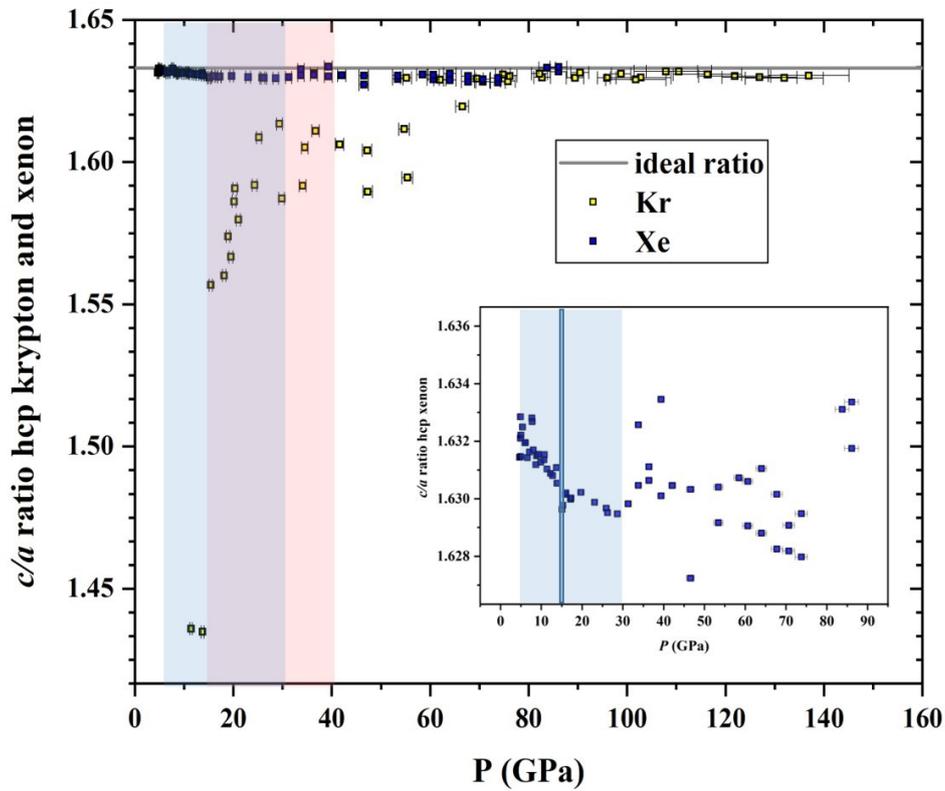

**Figure S7.** Pressure dependency of the *c/a* ratio for hcp xenon (blue squares) obtained in run-1 compared to those obtained for Kr in our previous work (yellow squares) [8]. **Uncertainties on the *c/a* are smaller than the symbol sizes. The ideal *c/a* ratio for close-packed structures corresponds to the sqrt(8/3) = 1.633. Note, that the *c/a* ratio of xenon is nearly ideal and that its variation with pressure is small considering that the axial ratios of most of the hcp metals fall in the range 1.57 < c/a > 1.65 [9]. As discussed in Kenichi [9] also under compression the *c/a* ratio shows very little variations for most hcp metals. For Kr a drastic variation of this ratio is observed at low pressure where the hcp fraction is very small and below 0.05 %.**

**Table SIV.** Pressure evolution of interatomic distance $R_i$ (also often refer to as the half-path lengths) for 4 single-scattering paths and their variations $\sigma^2_i$ also often referred to as the mean-square relative displacement Debye-Waller factor) of fcc and hcp xenon (run-3) extracted from the EXAFS fit. The uncertainties on pressure, evaluated from the pressure differences before and after an acquisition, measured with the ruby luminescence gauge [3] is smaller than 0.5% up to 90 GPa. The intrinsic uncertainties of the ruby pressure gauges are 2%. Above 90 GPa the pressure was determined from diffraction on the Re gasket. The $R$-factor is listed for each fit and presents a measure of the misfit between the raw data and the adjusted spectrum in % (a $R$-factor of 0.02 indicates a 2% misfit to raw data). At low pressure up to 10 GPa fits contained 20 independent number of points and 13 adjusted parameters comprising 6 paths with each individually fitted parameters $R$ and $\sigma^2$ and the energy shift $\Delta E$ and 8 background parameters. The fit window was set between 2 and 5.7 Å. Above this pressure the fits contained 60 independent number of points and 17 adjusted structural parameters comprising 8 paths with each individually fitted parameters $R$ and $\sigma$ and the energy shift $\Delta E$ and 16 fitted background parameters. The fit window was set between 2 and 8.5 Å. For all fits the amplitude reduction factor was set to 1. The relative uncertainty on first-neighbour distance ($R_1$) is smaller than ($5*10^{-3}$) over the entire pressure range. The error on the extracted bond distance variation are in the order of 3%. Variations of $\Delta E$ may arise from the growing hcp phase that modifies the white line as shown in [8] for krypton.

| Xenon (fcc) | | | $R_1$ (Å) N=12 | | | $R_2$ (Å) N=6 | | | $R_3$ (Å) N=24 | | | $R_5$ (Å) N=24 | | |
| P (GPa) | R-factor | $\Delta E$ | XRD | EXAFS | $\sigma^2_1$ (Å²) (*10^-3) | XRD | EXAFS | $\sigma^2_2$ (Å²) (*10^-3) | XRD | EXAFS | $\sigma^2_3$ (Å²) (*10^-3) | XRD | EXAFS | $\sigma^2_5$ (Å²) (*10^-3) |
|---|---|---|---|---|---|---|---|---|---|---|---|---|---|---|
| 5.9(4) | 0.02 | 0.9(1.1) | 3.836 | 3.786(21) | 15.2(1.6) | 5.424 | 5.37(17) | 27(30) | 6.64 | | | 8.58 | | |
| 8.5(4) | 0.044 | 0.6(1.7) | 3.745 | 3.701(30) | 12.8(2.0) | 5.297 | 5.22(24) | 23(53) | 6.49 | | | 8.38 | | |
| 10.5(6) | 0.057 | 0.1(1.2) | 3.691 | 3.676(18) | 11.8(48) | 5.220 | 5.19(13) | 13(24) | 6.39 | 6.38(8) | 14.3(9.8) | 8.25 | | |
| 15.3(2) | 0.037 | 1.1(0.8) | 3.589 | 3.563(11) | 9.0(8) | 5.076 | 5.02(9) | 9(13) | 6.22 | 6.20(12) | 17(13) | 8.03 | 8.09(12) | 13.5(14.4) |
| 16.1(1) | 0.03 | 0.2(0.6) | 3.575 | 3.553(6) | 9.2(5) | 5.056 | 4.99(4) | 8.7(5.0) | 6.19 | 6.18(3) | 13.0(3.3) | 8.00 | 7.98(7) | 13.0(7.5) |
| 23.8(6) | 0.029 | -0.4(0.7) | 3.465 | 3.43(6) | 6.6(4) | 4.900 | 4.84(5) | 6.1(5.2) | 6.00 | 5.96(4) | 10.2(6.4) | 7.75 | 7.77(1) | 9.1(8.0) |
| 30.3(4) | 0.028 | -0.5(6) | 3.394 | 3.375(5) | 5.5(4) | 4.800 | 4.75(5) | 6(6) | 5.88 | 5.86(4) | 10.1(3.8) | 7.59 | 7.62(7) | 11.0(8.2) |
| 38.3(8) | 0.025 | -0.2(6) | 3.325 | 3.306(5) | 4.8(3) | 4.702 | 4.67(6) | 6.4(6.8) | 5.76 | 5.75(4) | 7.3(4.2) | 7.44 | 7.48(6) | 8.5(6.0) |
| 53.1(1.0) | 0.02 | 0.8(5) | 3.226 | 3.222(3) | 4.5(2) | 4.562 | 4.57(3) | 6.5(2.3) | 5.59 | 5.59(2) | 7.5(1.6) | 7.21 | 7.26(3) | 7.7(2.7) |
| Xenon (hcp) | | | | | | | | | | | | | | |
| 15.3(2) | 0.03 | 0.2(8) | 3.574 | 3.547(5) | 9.2(6) | 5.055 | 4.95(2) | 3.8(4.5) | 6.19 | 6.23(3) | 2.1(4.0) | 7.99 | 7.94(5) | 5.2(6.2) |
| 16.1(1) | 0.03 | 0.2(3) | 3.560 | 3.552(5) | 9.2(4) | 5.035 | 5.03(2) | 8.3(2.0) | 6.17 | 6.20(2) | 11.9(3.6) | 7.96 | 7.97(6) | 8.9(6.4) |
| 23.8(6) | 0.036 | -0.5(6) | 3.450 | 3.430(5) | 6.7(4) | 4.879 | 4.84(4) | 7.4(4.8) | 5.98 | 5.99(4) | 10.0(4.2) | 7.72 | 7.75(5) | 6.0(5.2) |
| 30.3(4) | 0.022 | -0.7(4) | 3.380 | 3.375(4) | 5.8(3) | 4.781 | 4.76(4) | 5.9(4.3) | 5.86 | 5.90(2) | 9.2(2.4) | 7.56 | 7.65(6) | 7.2(5.8) |
| 38.3(8) | 0.024 | -0.2(8) | 3.312 | 3.306(6) | 4.8(3) | 4.683 | 4.64(4) | 3(3.6) | 5.74 | 5.80(4) | 8.0(4.2) | 7.41 | 7.49(7) | 6.2(6.8) |
| 53.1(1.0) | 0.023 | 0.5(3) | 3.214 | 3.221(3) | 4.5(2) | 4.545 | 4.55(4) | 7(15) | 5.57 | 5.61(2) | 7.7(1.4) | 7.19 | 7.25(3) | 4.9(2.7) |
| 61.33 | 0.018 | 0.4(5) | 3.170 | 3.200(3) | 4.4(2) | 4.483 | 4.53(2) | 5.(2) | 5.49 | 5.57(2) | 6.0(1.4) | 7.09 | 7.23(7) | 8.9(6.6) |
| 69.9(2.6) | 0.015 | 0.3(5) | 3.130 | 3.154(3) | 3.8(2) | 4.427 | 4.48(3) | 6.3(2.3) | 5.42 | 5.49(1) | 6.8(1.2) | 7.00 | 7.09(1) | 2.7(6.0) |
| 76.4(1) | 0.012 | 0.2(5) | 3.103 | 3.130(2) | 3.8(1) | 4.388 | 4.43(2) | 5.2(1.6) | 5.38 | 5.45(3) | 6.8(1.1) | 6.94 | 7.04(1) | 7.2(4.5) |
| 85.9(2.2) | 0.013 | 0.2(4) | 3.067 | 3.098(2) | 3.4(1) | 4.338 | 4.40(1) | 4.6(1.3) | 5.31 | 5.38(1) | 6.0(9) | 6.86 | 7.00(2) | 4.1(1.7) |
| 85.2(3) | 0.011 | 0.3(4) | 3.063 | 3.096(2) | 3.4(1) | 4.332 | 4.39(2) | 5.6(1.6) | 5.31 | 5.38(1) | 6.2(9) | 6.85 | 6.96(4) | 5.3(4.8) |
| 89.7(6) | 0.011 | 0.2(4) | 3.051 | 3.082(2) | 3.4(1) | 4.315 | 4.36(2) | 5.9(1.8) | 5.28 | 5.36(1) | 6.0(8) | 6.82 | 6.96(2) | 4.6(1.9) |
| 96.8(2) | 0.009 | 1.3(4) | 3.041 | 3.050(2) | 3.2(1) | 4.300 | 4.33(2) | 3.3(1.0) | 5.27 | 5.33(2) | 5.1(1.4) | 6.80 | 6.87(2) | 3.4(1.6) |
| 104.4(5) | 0.011 | 0.8(4) | 3.016 | 3.023(2) | 2.9(1) | 4.265 | 4.27(2) | 5.0(1.6) | 5.22 | 5.26(1) | 5.1(7) | 6.74 | 6.79(2) | 3.9(1.7) |
| 110(2) | 0.011 | 1.3(4) | 2.999 | 3.009(2) | 2.9(1) | 4.241 | 4.26(3) | 5.8(2.1) | 5.19 | 5.23(1) | 5.3(8) | 6.71 | 6.77(2) | 3.4(1.5) |
| 119(1) | 0.009 | 1.3(4) | 2.974 | 2.991(2) | 3.0(1) | 4.206 | 4.23(2) | 5.6(1.5) | 5.15 | 5.20(2) | 5.7(8) | 6.65 | 6.72(2) | 3.7(1.3) |
| 118(1) | 0.013 | 1.3(4) | 2.976 | 2.986(2) | 2.9(1) | 4.208 | 4.25(2) | 5.0(1.6) | 5.15 | 5.19(1) | 5.1(7) | 6.65 | 6.71(2) | 3.7(1.8) |
| 124(1) | 0.008 | 1.2(4) | 2.961 | 2.973(2) | 2.5(1) | 4.187 | 4.22(2) | 4.9(1.4) | 5.13 | 5.17(1) | 4.5(6) | 6.62 | 6.68(1) | 2.9(1.2) |
| 132(4) | 0.011 | 1.0(4) | 2.940 | 2.962(2) | 2.5(1) | 4.158 | 4.21(2) | 4.7(1.6) | 5.09 | 5.15(1) | 4.5(6) | 6.57 | 6.66(2) | 2.8(1.4) |
| 134(1) | 0.009 | 0.5(4) | 2.923 | 2.947(2) | 2.7(1) | 4.133 | 4.20(2) | 6.1(1.4) | 5.06 | 5.12(1) | 4.8(6) | 6.54 | 6.63(2) | 4.2(1.7) |
| 143(2) | 0.008 | 1.4(4) | 2.916 | 2.941(2) | 2.7(1) | 4.124 | 4.16(2) | 5.2(1.5) | 5.05 | 5.11(1) | 4.6(5) | 6.52 | 6.60(1) | 3.1(1.2) |
| 145(3) | 0.012 | 2.3(5) | 2.910 | 2.939(2) | 2.7(1) | 4.116 | 4.18(2) | 5.1(1.9) | 5.04 | 5.11(1) | 4.5(6) | 6.51 | 6.61(2) | 3.7(1.8) |
| 148(2) | 0.009 | 2.1(4) | 2.905 | 2.93(2) | 2.7(1) | 4.108 | 4.17(2) | 5.3(1.7) | 5.03 | 5.09(1) | 4.6(5) | 6.50 | 6.58(1) | 3.7(1.5) |
| 158.1(3) | 0.011 | 2.4(5) | 2.883 | 2.921(2) | 2.8(1) | 4.078 | 4.16(2) | 6.3(3.1) | 4.99 | 5.07(1) | 4.7(6) | 6.45 | 6.56(2) | 3.0(1.3) |

**Table SV.** Pressure evolution of interatomic distance $R_i$ (also often refer to as the half-path lengths) for 4 single-scattering paths and their variations $\sigma^2_i$, also often referred to as the mean-square relative displacement Debye-Waller factor) of fcc krypton (run-4 and -5) extracted from the EXAFS fit. The uncertainties on pressure, evaluated from the pressure differences before and after an acquisition, measured with the ruby luminescence gauge [3], is smaller than 0.5% up to 100 GPa. The intrinsic uncertainties of the ruby pressure gauges are 2%. The $R$-factor is listed for each fit and presents a measure of the misfit between the raw data and the adjusted spectrum in % (a $R$-factor of 0.02 indicates a 2% misfit to raw data). Each fit contained 35 independent number of points and 17 adjusted parameters: 8 paths with each comprised each individually fitted parameters $R$ and $\sigma^2$ and the energy shift $\Delta E$. The amplitude reduction factor was set to 1. The relative uncertainty on first-neighbour distance ($R_1$) is smaller than $(5*10^{-3})$ over the entire pressure range. The error on the extracted bond distance variation are in the order of 3%. Variations of $\Delta E$ may arise from the growing hcp phase that modifies the white line as shown in [8].

| **Krypton (fcc)** | | | $R_1$ (Å) N=12 | | | $R_2$ (Å) N=6 | | | $R_3$ (Å) N=24 | | | $R_5$ (Å) N=24 | | |
|---|---|---|---|---|---|---|---|---|---|---|---|---|---|---|
| $P$ (GPa) | R-factor | $\Delta E$ | XRD | EXAFS | $\sigma^2_1$ (Å²) (*10^-3) | XRD | EXAFS | $\sigma^2_2$ (Å²) (*10^-3) | XRD | EXAFS | $\sigma^2_3$ (Å²) (*10^-3) | XRD | EXAFS | $\sigma^2_5$ (Å²) (*10^-3) |
| 15.4(2) | 0.026 | 1.3(2) | 3.288 | 3.302(3) | 13.8(4) | 4.649 | 4.71(2) | 15(3) | 5.694 | 5.785(11) | 17(8) | 7.351 | 7.419(1) | 20(4) |
| 18.1(2) | 0.024 | -0.8(3) | 3.248 | 3.221(4) | 13.2(5) | 4.593 | 4.65(2) | 14(3) | 5.626 | 5.656(44) | 29(7) | 7.263 | 7.35(4) | 17(5) |
| 21.3(3) | 0.031 | -0.2(3) | 3.208 | 3.187(3) | 11.6(4) | 4.536 | 4.67(4) | 18(5) | 5.556 | 6.203(13) | 13(2) | 7.172 | 7.00(7) | 25(10) |
| 23.1(3) | 0.020 | -1.4(3) | 3.188 | 3.143(3) | 9.5(3) | 4.508 | 4.54(4) | 18(6) | 5.521 | 5.449(7) | 10(1) | 7.128 | 7.01(4) | 23(10) |
| 26.6(3) | 0.010 | -1.6(2) | 3.153 | 3.114(2) | 8.4(2) | 4.458 | 4.11(9) | 18(14) | 5.460 | 5.417(4) | 10(1) | 7.049 | 6.96(7) | 27(10) |
| 28.6(3) | 0.018 | -1.5(2) | 3.134 | 3.077(2) | 7.4(2) | 4.432 | 4.36(2) | 88(1) | 5.428 | 5.357(4) | 9.4(5) | 7.008 | 6.90(6) | 25(8) |
| 32.3(4) | 0.006 | -1.4(1) | 3.104 | 3.071(4) | 7.3(2) | 4.390 | 4.34(2) | 26(14) | 5.377 | 5.361(10) | 9(1) | 6.941 | 6.88(4) | 22(5) |
| 38.0(5) | 0.018 | 0.5(3) | 3.064 | 3.070(2) | 6.5(3) | 4.333 | 4.35(8) | 5(1) | 5.307 | 5.381(12) | 13(1) | 6.851 | 6.86(5) | 18(6) |
| 39.8(5) | 0.019 | 0.6(3) | 3.053 | 3.064(2) | 6.1(3) | 4.317 | 4.34(8) | 4(1) | 5.288 | 5.369(15) | 14(2) | 6.826 | 6.85(4) | 16(5) |
| 40.6(5) | 0.017 | 0.6(3) | 3.048 | 3.060(2) | 6.1(3) | 4.310 | 4.34(8) | 4(1) | 5.279 | 5.364(16) | 15(2) | 6.815 | 6.85(4) | 16(5) |
| 41.7(5) | 0.022 | 0.4(3) | 3.041 | 3.053(3) | 5.8(3) | 4.301 | 4.33(9) | 4(1) | 5.268 | 5.329(13) | 12(2) | 6.801 | 6.83(3) | 13(4) |
| 42.7(5) | 0.024 | 0.2(3) | 3.035 | 3.046(3) | 5.7(3) | 4.293 | 4.32(9) | 4(1) | 5.257 | 5.295(9) | 9(1) | 6.787 | 6.82(3) | 13(4) |
| 43.8(5) | 0.027 | -0.1(3) | 3.029 | 3.035(3) | 5.6(3) | 4.284 | 4.32(1) | 5(1) | 5.247 | 5.258(7) | 7(1) | 6.774 | 6.80(3) | 12(4) |
| 45.3(5) | 0.021 | -0.2(3) | 3.021 | 3.025(3) | 5.4(3) | 4.272 | 4.31(1) | 5(1) | 5.232 | 5.247(6) | 7(1) | 6.755 | 6.79(3) | 11(3) |
| 47.4(6) | 0.026 | -0.5(3) | 3.010 | 3.008(3) | 5.2(3) | 4.256 | 4.29(2) | 5(2) | 5.213 | 5.208(5) | 7(1) | 6.730 | 6.76(3) | 10(3) |
| 49.5(6) | 0.021 | -0.8(3) | 2.999 | 2.994(3) | 5.0(3) | 4.242 | 4.27(2) | 8(2) | 5.195 | 5.184(5) | 5(1) | 6.706 | 6.75(3) | 10(2) |
| 52.4(6) | 0.020 | -1.1(3) | 2.986 | 2.982(2) | 4.8(3) | 4.223 | 4.24(4) | 4(6) | 5.172 | 5.164(4) | 4.0(5) | 6.677 | 6.73(3) | 11(3) |
| 55.1(7) | 0.015 | -1.0(2) | 2.973 | 2.961(2) | 4.2(2) | 4.205 | 4.19(2) | 9(2) | 5.150 | 5.137(4) | 4.3(4) | 6.649 | 6.73(3) | 13(4) |
| 56.9(7) | 0.012 | -0.7(2) | 2.966 | 2.955(2) | 4.1(2) | 4.194 | 4.19(1) | 8(2) | 5.137 | 5.132(4) | 5.3(5) | 6.631 | 6.73(3) | 13(3) |
| 58.5(7) | 0.011 | -0.9(2) | 2.959 | 2.949(2) | 4.1(2) | 4.185 | 4.17(1) | 8(2) | 5.125 | 5.115(4) | 4.9(4) | 6.616 | 6.73(3) | 14(4) |
| 61.0(7) | 0.008 | -0.9(2) | 2.949 | 2.939(1) | 3.9(2) | 4.171 | 4.16(1) | 8(1) | 5.108 | 5.100(4) | 5.5(4) | 6.594 | 6.73(3) | 13(3) |
| 64.9(8) | 0.010 | -0.8(2) | 2.934 | 2.926(2) | 3.8(2) | 4.150 | 4.13(1) | 8(1) | 5.082 | 5.069(4) | 6(1) | 6.561 | 6.72(3) | 14(4) |
| 67.6(8) | 0.010 | -0.7(2) | 2.925 | 2.923(1) | 3.5(2) | 4.136 | 4.13(1) | 7(1) | 5.066 | 5.072(5) | 6(1) | 6.540 | 6.71(4) | 15(5) |

**A**

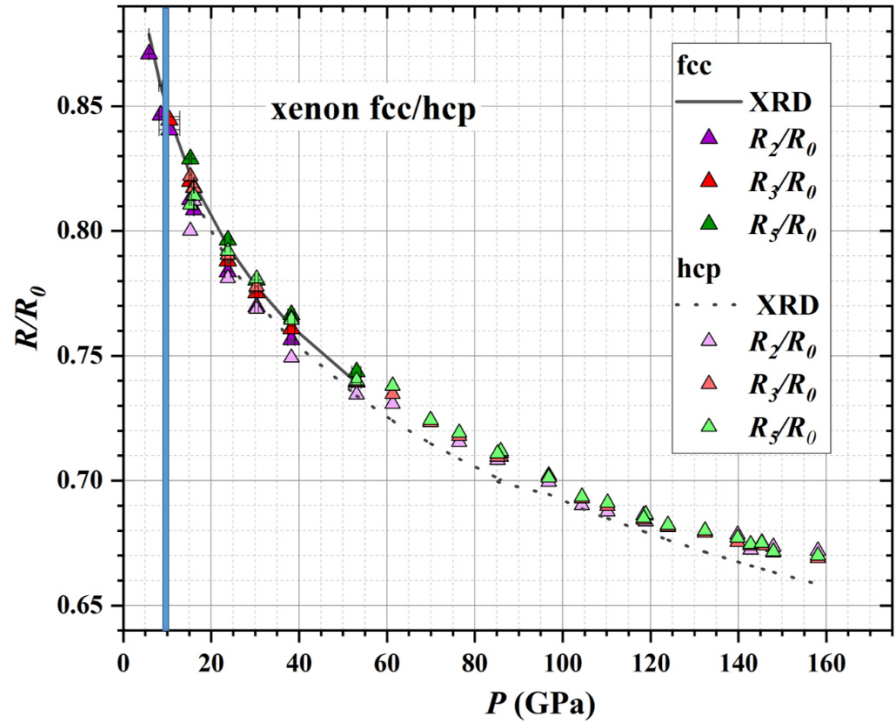

**B**

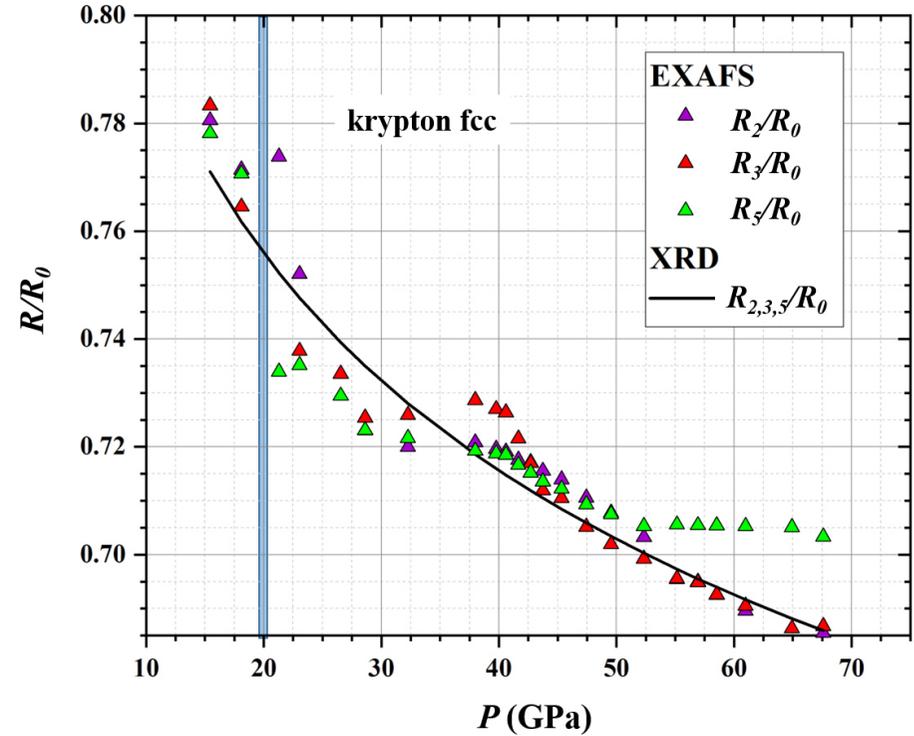

**Figure S8.** Pressure evolution of the normalized EXAFS fitted half-path length ($R/R_0$) for the $2^{nd}$, $3^{rd}$ and $5^{th}$ path in fcc or hcp xenon (**A**) and fcc krypton (**B**) plotted together with those calculated form the EoS of fcc krypton derived from XRD data from [8] and xenon derived from XRD data of the present study. Uncertainties on $R/R_0$ and pressure are smaller than the symbol size. The blue bar indicates the pressure of the compression anomaly seen from XRD. Symbol colours correspond to those of Fig. 6 in the main text. Note, that in the pressure interval right after the compression anomaly (20-30 GPa for krypton and (15-35 GPa for xenon) fitted scattering paths lengths show systematically a negative deviation from the ideal fcc value calculated from diffraction. This can be explained by the growing hcp stacking faults that exhibits a lower lattice volume compared to the hcp phase. Beyond 40 GPa in both systems a good agreement to XRD data is observed. Above 50 GPa the data suggest a positive deviation from the ideal value. Note the deviation from ideal behaviour of the next nearest neighbour variances right 15-20 GPa after the compression anomaly, its conversion at higher pressures (40-50 GPa for krypton and xenon) and its subsequent divergence beyond 50 GPa in both systems.

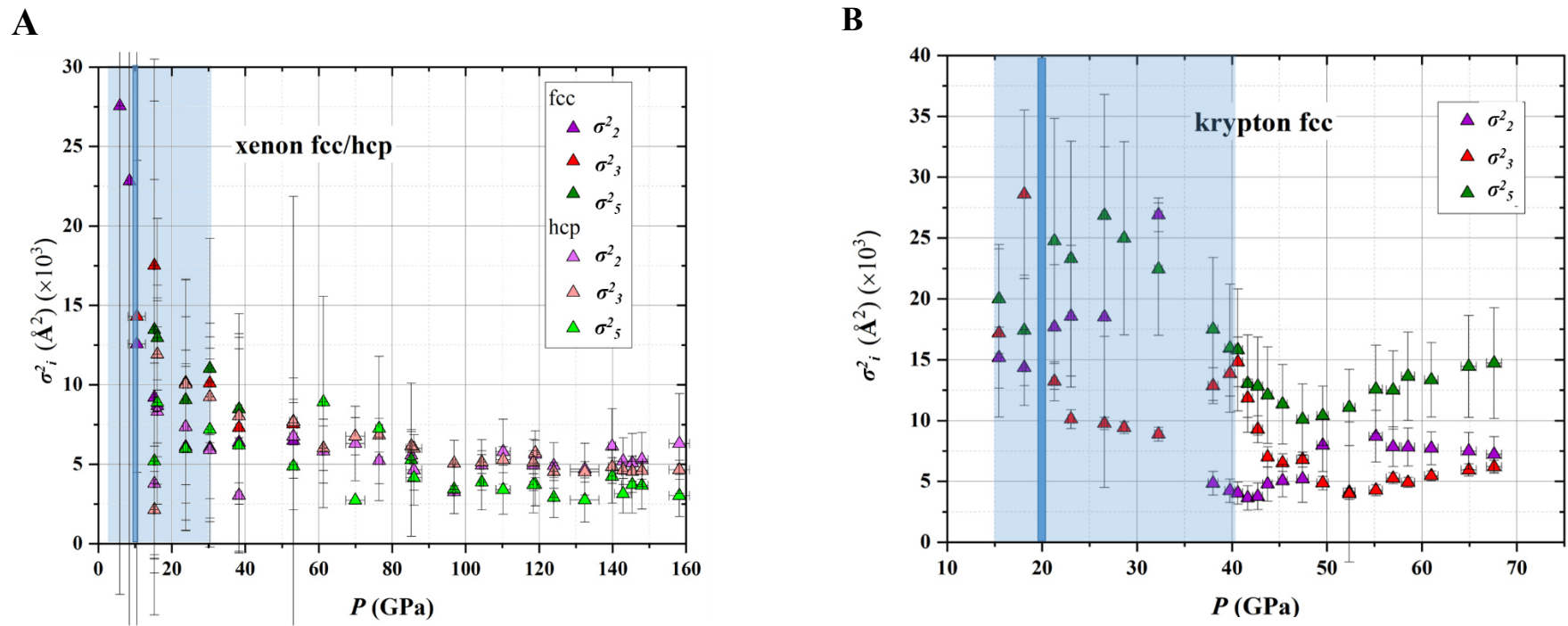

**Figure S9.** Pressure dependence of the fitted next-nearest neighbor distributions ($\sigma^2_i$) for the 2nd, 3rd and 5th neighbours in fcc or hcp xenon (**A**) and fcc krypton (**B**). The blue bar indicates the pressure of the compression anomaly seen from XRD. Symbol colours correspond to those of Fig. 6 in the main text. Note the high variation of this value 15-20 GPa after the anomaly and their convergence at higher pressures.